# Freestanding n-Doped Graphene *via* Intercalation of Calcium and Magnesium Into the Buffer Layer – SiC(0001) Interface

*Jimmy C. Kotsakidis[1*], Antonija Grubišić-Čabo[1], Yuefeng Yin[2], Anton Tadich[3], Rachael L. Myers-Ward[4], Matthew DeJarld[4], Shojan P. Pavunny[4], Marc Currie[4], Kevin M. Daniels[5], Chang Liu[1], Mark T. Edmonds[1], Nikhil V. Medhekar[2], D. Kurt Gaskill[5], Amadeo L. Vázquez de Parga[6,7*], Michael S. Fuhrer[1*]*

[1] School of Physics and Astronomy and [2] Department of Materials Science and Engineering, Monash University, *Melbourne, Victoria 3800, Australia.*

[3] Australian Synchrotron, *800 Blackburn Rd, Melbourne, Victoria 3168, Australia.*

[4] U.S. Naval Research Laboratory, *Washington D.C. 20375, USA.*

[5] Institute for Research in Electronics and Applied Physics, University of Maryland, *College Park, MD, 20742, USA*

[6] Dep. Física de la Materia Condensada and Condensed Matter Physics Center (IFIMAC), Universidad *Autónoma de Madrid, Cantoblanco 28049, Madrid, Spain.*

[7] IMDEA *Nanociencia, Cantoblanco 28049, Madrid, Spain.*

*Address correspondence to jimmy.kotsakidis@monash.edu, michael.fuhrer@monash.edu or al.vazquezdeparga@uam.es

**Abstract**

The intercalation of epitaxial graphene on SiC(0001) with Ca has been studied extensively, yet precisely where the Ca resides remains elusive. Furthermore, the intercalation of Mg underneath epitaxial graphene on SiC(0001) has not been reported. Here, we use low energy electron diffraction, x-ray photoelectron spectroscopy, secondary electron cut-off photoemission and scanning tunneling microscopy to elucidate the physical and electronic structure of both Ca- and Mg-intercalated epitaxial graphene on 6H-SiC(0001). We find that Ca intercalates underneath the buffer layer and bonds to the Si-terminated SiC surface, breaking the C-Si bonds of the buffer layer *i.e.* 'freestanding' the buffer layer to form Ca-intercalated quasi-freestanding bilayer graphene (Ca-QFSBLG). The situation is similar for the Mg-intercalation of epitaxial graphene on SiC(0001), where an ordered Mg-terminated reconstruction at the SiC surface is formed and Mg bonds to the Si-terminated SiC surface are found, resulting in Mg-intercalated quasi-freestanding bilayer graphene (Mg-QFSBLG). Ca-intercalation underneath the buffer layer has not been considered in previous studies of Ca-intercalated epitaxial graphene. Furthermore, we find no evidence that either Ca or Mg intercalates between graphene layers. However, we do find that both Ca-QFSBLG and Mg-QFSBLG exhibit very low workfunctions of 3.68 and 3.78 eV, respectively, indicating high n-type doping. Upon exposure to ambient conditions, we find Ca-QFSBLG degrades rapidly, whereas Mg-QFSBLG remains remarkably stable.

Graphene, the two dimensional (2D) allotrope of carbon consisting of a network of hexagonally bonded carbon atoms,[1] possesses radically different properties than that of its bulk form (graphite), owing to its Dirac cone band structure at low energy.[2] In addition, considerable research has been conducted on the modification of graphene's properties *via* doping using techniques such as electrostatic gating,[3] surface decoration,[4-5] structure modification[6] and intercalation.[7] Of these methods, intercalation (*i.e.*, insertion of atomic or molecular species under or in-between graphene layers) has proven a powerful method for achieving the highest doping levels in graphene.[8-10] Thus, intercalation has enabled various investigations into the fundamental physics of superconductivity[8, 11-12] and many-body interactions[8-10] as well as potential applications such as highly conductive and transparent electrodes[13] and energy storage.[14]

Following the discovery of superconducting Ca-intercalated graphite ($CaC_6$) – which exhibits the highest superconducting transition temperature ($T_c$ = 11.5 K) amongst the graphite intercalation compounds – intercalation of the alkaline-earth Ca has been of particular interest in attempts to achieve superconductivity[12] and extremely high doping in graphene.[8] Superconductivity with a high-$T_c$ *via* conventional electron-phonon coupling[15] or unconventional electron-electron coupling[16-17] has been predicted in highly n-type doped graphene if the Fermi level ($E_F$) is shifted high into the conduction band near the Van Hove singularity. This scenario has analogies to the high-$T_c$ cuprates, in which superconductivity is associated with the interaction between $E_F$ and the Van Hove singularity.[18] This physical scenario in graphene was first observed by McChesney *et al.*[8] using sub-monolayer graphene on SiC(0001) intercalated with Ca (and the surface decorated with Ca and K), and it was subsequently suggested that the sample could exhibit superconductivity.

More recently, the superconductivity of Ca-intercalated bilayer graphene was reported by Ichinokura et al.[12] with $T_c \approx 2K$.

The Ca-intercalation experiments performed to date have utilized large-area graphene on SiC,[8, 11-12, 19-21] as it enables a wide range of surface probes such as scanning tunneling microscopy (STM), low energy electron diffraction (LEED), X-ray photoelectron spectroscopy (XPS) and angle resolved photoemission spectroscopy (ARPES). Graphene synthesis on SiC occurs *via* the sublimation of Si, where the remaining C naturally forms a hexagonal carbon network. This first layer of carbon is often termed the "zero layer" or "buffer layer",[22] and is partially bonded to the topmost silicon on SiC(0001). Continued sublimation forms another buffer layer beneath the first, releasing the C-Si bonds of the original buffer layer, which then becomes monolayer graphene.

The structure of Ca-intercalated graphene on SiC is less clear, with disagreement regarding the location of the intercalated Ca. For instance, McChesney et al.[8] implied the position of the Ca intercalant in their partial monolayer graphene was between the 1st graphene layer and the buffer layer. Kanetani et al.[19] and Ichinokura et al.[12] studied Ca-intercalated bilayer graphene on SiC(0001), and based on LEED and reflection high energy electron diffraction data, concluded that Ca was intercalated between the 1st and 2nd graphene layers. A correction to this interpretation was recently published by Endo et al.[20], in which it was determined that the Ca intercalant position for bilayer graphene on SiC was between the buffer layer and 1st graphene layer.

The calcium intercalation of graphite [23-24] and graphene [8, 12, 19-21, 25-26] has been extensively studied, and can highly n-type dope these systems.[8] On the other hand, the alkaline-earth magnesium has not been investigated as an intercalant in graphene on silicon carbide systems. Some reasons for this may be due to the well-known experimental knowledge that metallic magnesium does not intercalate graphite,[27] and also may be exacerbated by magnesium's well-known high vapour pressure.[28] More recent theoretical considerations have substantiated the difficulty of magnesium intercalation for a range of different materials (including graphite), and have implied that the difficulty of Mg intercalation is a result of its weak binding forces to the target substrate.[29] Thus, it is likely that for these reasons that the intercalation of graphene on SiC with magnesium was not considered.

Despite these challenges, magnesium has been intercalated underneath graphene on Ni(111),[30] and between layers of a 'graphite-like' material composed of boron carbon and nitrogen.[31] Furthermore, magnesium intercalation for energy storage applications has long been sought after to replace lithium.[32] But in this case, magnesium forms part of an electrolyte or organic molecule.[33-34] As a rule, it has generally been understood that Mg-intercalation of pure carbon systems is not possible, but the fact that intercalation of graphene on SiC(0001) systems can occur at the buffer layer – SiC(0001) interface[35] implies that the 'traditional' rules of intercalation from experiments with graphite in which the intercalant is inserted between graphite layers, may not apply (as was the case with other substrates and modified materials[30-31]).

In this communication, we present new findings on the intercalation structure of two alkaline-earth metals – Ca and Mg – *via* the intercalation of epitaxial monolayer graphene (EMLG) (and hydrogen intercalated 'quasi-freestanding' bilayer graphene, 'H-QFSBLG' – see Supporting Information, Section 1.3) on 6H-SiC(0001). We show that Ca intercalates underneath the buffer layer of EMLG, bonding with the Si-terminated surface of the SiC(0001) and 'quasi-freestanding' the buffer layer to create another graphene layer, *i.e.*, forming quasi-freestanding bilayer graphene (QFSBLG) above the Ca-Si layer, which we term Ca-QFSBLG. The intercalated Ca forms a new ordered (√3×√3)R30° interface reconstruction with respect to the graphene. The observations are supported by density functional theory (DFT) calculations, which show Ca under the buffer layer as the lowest energy configuration. In addition, we report for the first time on the intercalation of EMLG with Mg. Similar to Ca, Mg intercalates underneath the buffer layer of EMLG, bonding with the Si surface of the SiC and freestanding the buffer layer to again form QFSBLG, which we term Mg-QFSBLG. Here, stronger Mg-Si interactions produce an ordered Mg-terminated (√3×√3)R30° reconstruction with respect to the SiC rather than the graphene. As in the Ca-intercalation experiment, we find no evidence to suggest the formation of an Mg-carbide or intercalation of the Mg in-between graphene layers. Both Ca- and Mg-intercalated systems result in the formation of bilayer graphene with exceptionally low workfunctions (3.68 and 3.78 eV, respectively). However, while the Ca-QFSBLG system is not air-stable, we find that the Mg-QFSBLG system shows remarkable stability to ambient conditions, with little change in photoemission spectra after ≈ 6 hours of exposure to ambient conditions. Consequently, Mg-intercalation could offer a possible route to highly electron doped, low work function and air-stable graphene.

## Experimental Overview

The EMLG samples were prepared using a previously described method (further details can be found in the Methods).[36] The EMLG samples are majority monolayer, with inclusions of bilayer and trilayer regions. However, our conclusions are not affected by the number of graphene layers in the starting sample, as (1) the LEED and XPS analysis of the structure and chemistry at the intercalated buffer layer/SiC interface is insensitive to the number of graphene overlayers, and (2) previous studies have shown that bilayer regions (step edges) of such samples do not include a buffer layer,[37] and subsequent intercalation (with hydrogen) results in near-total conversion of the samples to QFSBLG.[38] Ca and Mg intercalation for XPS and LEED measurement was carried out at the Australian Synchrotron soft x-ray beamline end station under ultrahigh vacuum (UHV) conditions ($\approx 1 \times 10^{-10}$ mbar). Low temperature STM was carried out at Monash University on Ca-QFSBLG prepared *in situ*, under UHV ($\approx 1 \times 10^{-10}$ mbar). The Mg-QFSBLG sample analyzed with the STM was prepared by Mg intercalation at the Australian Synchrotron, and then transferred in ambient to the STM UHV chamber.

We have additionally studied Ca and Mg intercalation of H-QFSBLG, prepared using a previously described method (see Methods for details on growth).[38] We found no evidence for Mg intercalation of H-QFSBLG (see Supporting Information Section 2.3), but were able to intercalate Ca to transform H-QFSBLG to Ca-QFSBLG, with substantially similar properties to Ca-QFSBLG formed from EMLG (discussed further in Supporting Information Section 1.3). Details of all experiments are found in Methods.

## Results and Discussion

**Low Energy Electron Diffraction (LEED) of Ca- and Mg-intercalated EMLG on SiC(0001).**
Figure 1 shows LEED images taken at an electron energy of 100 eV. Fig. 1a shows a typical EMLG sample with graphene (G(1×1)) and SiC (SiC(1×1)) spots, as well as the expected (6√3×6√3)R30° and (6×6) spots which originate from the buffer layer – SiC interaction.[39] The Ca- and Mg-intercalated samples are shown in Fig. 1b and c, respectively, and share important features. Firstly, the relative intensity of the SiC(1×1) to the G(1×1) spots after Ca- and Mg-intercalation is greatly reduced. This is in contrast to the similar intensity of the G(1×1) and SiC(1×1) spots in pristine EMLG prior to intercalation (Fig. 1a). Secondly, the (6√3×6√3)R30° spots are strongly suppressed in both the Ca- and Mg-intercalated samples (Fig. 1b and c, respectively), when compared to pristine EMLG (Fig. 1a). These observations are similar to the case of H-QFSBLG after H-intercalation of EMLG, which is known to decouple the buffer layer,[40] and thus, suggests that the buffer layer of the EMLG is similarly decoupled after Ca/Mg-intercalation and transformed to QFSBLG (*i.e.* Ca/Mg-QFSBLG). Consequently, our LEED results present the first evidence that intercalation of Ca and Mg occurs between the SiC(0001) surface and the buffer layer, in disagreement with previous reports. We note that the weak (6√3×6√3)R30° spots after Ca/Mg-intercalation (seen in Figs. 1b and c), are also observed after H-intercalation,[7] and likely results from incomplete (*i.e.* partial) intercalation.[41]

Furthermore, Fig. 1b shows the emergence of (√3×√3)R30° spots with respect to the G(1×1) spots after Ca-intercalation (which we label 'G(√3×√3)R30°'). Previous Ca-intercalation studies[19] also reported G(√3×√3)R30° spots and interpreted these as due to Ca atoms intercalating between the graphene layers. Yet the G(√3×√3)R30° spots are not definitive evidence for intercalation between the graphene layers, as their emergence has been observed after the intercalation of Yb in buffer-layer-only samples to form 'quasi-freestanding' monolayer graphene on SiC.[10] Thus, it is plausible that the G(√3×√3)R30° spots instead describe a Ca-QFSBLG structure. As we discuss further below, the prior interpretation is incorrect, and Ca (and Mg) indeed lie between the SiC(0001) surface and the buffer layer.

In contrast, the Mg-intercalated EMLG sample in Fig. 1c shows the emergence of (√3×√3)R30° spots with respect to the SiC(1×1) spots, not the G(1×1) spots (i.e. SiC(√3×√3)R30°). Such spots were observed by Stöhr et al.[42] and Kim et al.[43] after intercalation of buffer layer-only on SiC with Bi and In, respectively, and by Xia et al.[44] after Si-intercalation of EMLG. The SiC(√3×√3)R30° spots have also been observed in the past with Si adatoms bonded *via* the dangling bonds of the Si surface of SiC.[45] Thus, the observation of these SiC(√3×√3)R30° spots strongly indicates that Mg has intercalated beneath the buffer layer, releasing the buffer layer – SiC bonds to create Mg-QFSBLG; in agreement with the disappearance of the (6√3×6√3)R30°/(6×6) and brightness increase in G(1×1) spots relative to the SiC(1×1) spots.

Also visible in Fig 1c are extra spots concentric around both the G(1×1) and SiC(1×1) spots. These spots are shown more clearly in the magnified and enhanced views in the right inset of Fig. 1c. In regards to the G(1×1) spots, the new concentric spots correspond to an (18×18) reconstruction, rotated 30° with respect to the (6√3×6√3)R30° spots. Observation of the (18×18) spots is consistent with the expected expansion of the Moiré pattern from the interaction of graphene with SiC(0001) from (6√3×6√3)R30° to (18×18) upon the (√3×√3)R30° reconstruction of the surface (*i.e.* (√3×√3)R30° × (6√3×6√3)R30° = (18×18)). These LEED spots have been observed previously after the intercalation of Pt underneath graphene on SiC(0001), and their appearance was attributed to the concurrent formation of a Pt-silicide.[46] Similarly, we observe the appearance of (6√3×6√3)R30° spots around the SiC(1×1) spots corresponding to the expansion of the (6×6) Moiré by the (√3×√3)R30° reconstruction. Also seen are (5×5) spots in Fig. 1c, but these were not found to have significance regarding the Mg-intercalation structure.[39]

LEED spot patterns have elucidated the overlaying symmetries resulting from Ca- and Mg-intercalation of EMLG and suggest that both Ca *and* Mg intercalate underneath the buffer layer to form Ca-QFSBLG and Mg-QFSBLG. Nonetheless, LEED alone cannot determine the chemistry occurring underneath the graphene, and so we turn to XPS. Below, we first discuss the Si 2p and then the C 1s XPS spectra of Ca-QFSBLG (discussion of the O 1s and Ca 2p core levels can be found in the Supporting Information, Section 1.5), before discussing the surface morphology of Ca-QFSBLG as imaged by STM, and density functional theory calculations concerning the predicted structure of Ca-intercalated EMLG. This discussion will then be repeated for Mg-QFSBLG (without the theoretical consideration), which we will show, shares many similarities with Ca-QFSBLG.

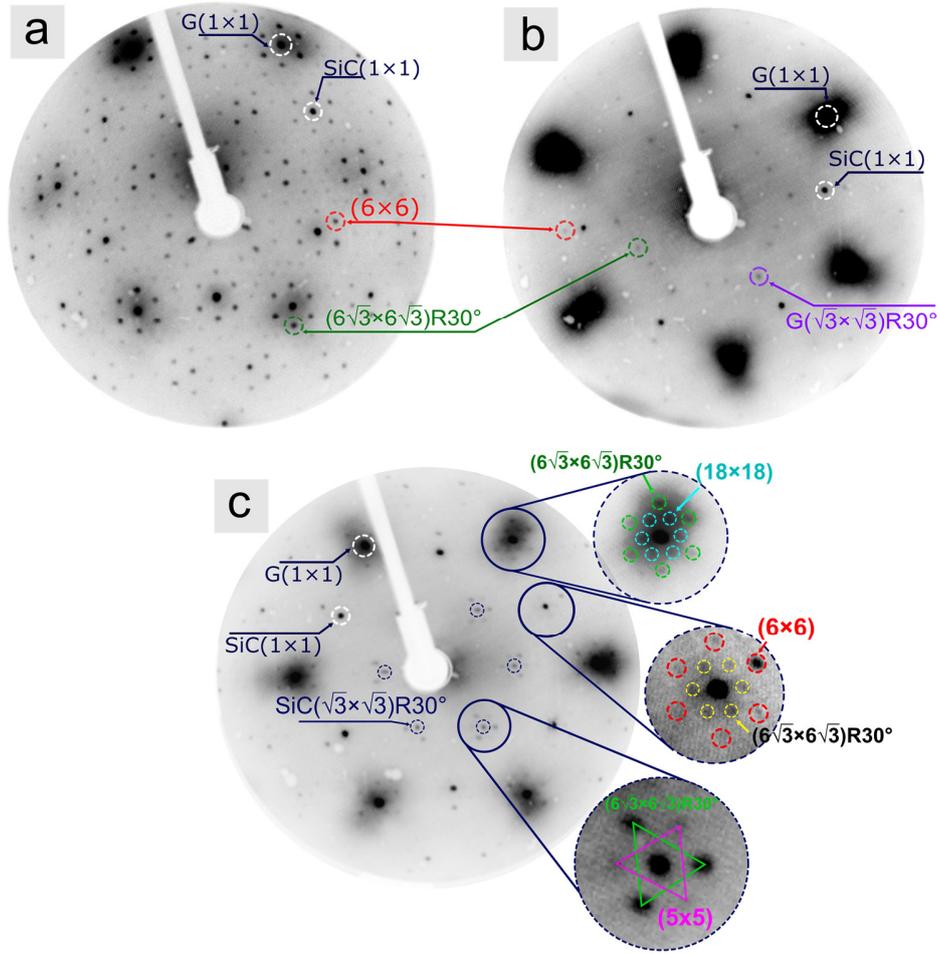

**Figure 1.** Low energy electron diffraction (LEED) patterns at 100 eV. **(a)** Pristine epitaxial monolayer graphene on SiC(0001) (EMLG) showing graphene (G(1×1)) and SiC (SiC(1×1)) spots, as well as (6√3×6√3)R30° and (6×6) spots arising from the buffer layer. **(b)** Ca-QFSBLG showing the emergence of (√3x√3)R30° spots with respect to the G(1×1) spots (G(√3x√3)R30°). **(c)** Mg-QFSBLG with spots corresponding to a (√3×√3)R30° reconstruction relative to the SiC(1×1) spots (SiC(√3x√3)R30°), along with (5×5), (18×18) (around the G(1×1)), and (6√3×6√3)R30° (around the SiC(1×1)) spots. In both Ca-QFSBLG (b) and Mg-QFSBLG (c) we observe the suppression of (6√3×6√3)R30°/(6×6) spots related to the buffer layer and increase in brightness of the G(1×1) with respect to the SiC(1×1) spots, as compared to pristine EMLG in (a).

**X-ray photoelectron spectroscopy of Ca-QFSBLG.** Here we present and discuss the XPS results for the C 1s and Si 2p core levels for the same EMLG sample, the LEED of which is shown in Fig. 1b. H-QFSBLG samples were also transformed to Ca-QFSBLG by the same intercalation method used for the EMLG samples, and showed similar XPS spectral features – see Supporting Information Section 1.3 for details.

Figure 2 shows the Si 2p core level XPS spectra – photoemission intensity as a function of binding energy, $E_B$ – before and after Ca-intercalation. In all XPS spectra, experimental data points are shown as open black circles, and the final fit is overlaid as a red line. Each Si 2p component is comprised of $2p_{3/2}$ and $2p_{1/2}$ peaks colored identically to avoid confusion, and we quote only $E_B$ values corresponding to the $2p_{3/2}$ peak. Fit parameters of the major components are given in Table 1; the full set of fit parameters (which includes the omitted 1st intercalation step) can be found in Supporting Information, Section 1.1.

Figs. 2a and 2b show the Si 2p core level spectra of pristine EMLG. The tunability of the synchrotron source allowed different x-ray energies ($E_{h\nu}$ = 150 eV, 600 eV) to be used in order to determine the surface and bulk nature of the various components, and aided in spectrum deconvolution. Pristine EMLG shows two major components: $B_{Si}$ ($E_B$ = 101.47 ± 0.05 eV) corresponds to "bulk" SiC,[7, 47-48] and $Z_{Si}$ ($E_B$ = 101.83 ± 0.05 eV) to surface Si bonded to the C in the buffer layer/zero layer.[7, 43, 49] Further discussion concerning component $Z_{Si}$ can be found in Supporting Information, Section 1.2. (The components corresponding to compounds present in both the Si 2p and C 1s XPS spectra are designated using the same label, with different subscript labels, in order to distinguish the specific core level, *i.e.*, $B_{Si}$ and $B_C$ are the bulk SiC signals in the Si 2p and C 1s spectra, respectively). We note here that the $E_B$ value was calculated by averaging all measured values (including the values from the omitted 1st intercalation which can be found in the Supporting Information, Section 1.1) – see Methods for further details. The corresponding atomic positions for the major components of pristine EMLG are shown schematically in the inset of Fig. 2b. These positions are approximately determined by a calculation of the relative intensity ratio, $RI_R$, discussed in detail in Methods. The $RI_R$ is a qualitative measure of surface sensitivity, with an $RI_R > 1$ indicating that the component is more towards the surface (see Methods and Supporting Information Section 1.1), and was calculated using the relative intensity (*RI*) referenced to the well-known bulk component $B_{Si}$ (or $B_C$ in the case of C 1s spectra, see below), following closely the conventions used in ref. [49]. For instance, the $RI_R$ of component $Z_{Si}$ is 1.25 [$RI(150\ eV)/RI(600\ eV) = 0.4/0.32 = 1.25$], and implies a more surface origin for this component. Additional components – $Ox_0$, $A_0$, $A_1$ and $A_2$ are seen with intensities <8% relative to component $B_{Si}$, and likely result from oxygen contamination ($Ox_0$) and silicon adatoms ($A_0$, $A_1$, $A_2$). These components, along with components $Ox_1$ and $Ox_2$ (which are thought to arise from differing Ca-Si stoichiometry), are in small concentration, and thus, are not significant for the determination of the general Ca-intercalated structure. We refer the reader to the Supporting Information Section 1.2 and 1.3 for further discussion of these features.

Figures 2c and 2d show the Si 2p XPS spectra of Ca-QFSBLG formed by Ca-intercalation of the sample in Fig. 2a and b. We show here only the final Ca-intercalation step. The results of the intermediate intercalation step, which do not affect the conclusions presented here, can be found in the Supporting Information, Section 1.2. Most of the bulk component, $B_{Si}$, has disappeared and has been replaced by several components at lower binding energy.

Component Ca$_{Si}$ is highly surface-like, with an $RI_R = 4$. It is substantially shifted relative to component B$_{Si}$ (and Z$_{Si}$) of pristine EMLG by $\Delta E_B = 2.18 \pm 0.11$ eV, suggesting a significant chemical change at the Si surface. Its binding energy of $E_B = 99.29 \pm 0.06$ eV, is similar to that of a Ca-silicide.[50-51] This is strong evidence for a Ca-Si bonding environment at the SiC surface underneath the graphene and buffer layer, and is in contrast with previous reports that Ca intercalates between the graphene layers,[12, 19] or between the buffer layer and the 1$^{st}$ graphene layer.[8, 20]

In fact, many previous reports on the intercalation of alkalis,[52-54] transition metals,[46, 55-56] rare earths[9-10, 57] and other elements[49, 58-60] have reported intercalation underneath the buffer layer (see ref. [35] for a brief review of intercalation of graphene on SiC). Furthermore, many prior studies on the intercalation of graphene on SiC have specifically reported the formation of silicide-like compounds,[46, 56, 60-61] and thus, it is not surprising that we observe a surface Si component matching closely in binding energy with that of a Ca-silicide. The observation of a Ca-Si interaction implies disruption of the Si-C bonds on the SiC(0001) surface with the C-rich buffer layer, and is supported by the LEED intensity suppression of the (6√3×6√3)R30° spots (which arise from the buffer layer), and an increase in the intensity of the G(1×1) spots relative to the SiC(1×1) spots which indicates decoupling of the buffer layer (see Fig. 1b). Both XPS and LEED observations are consistent with the formation of a Ca-Si compound and a new graphene layer.

Two other major peak components in Fig. 2c, d include components B$'_{Si}$ and Z$'_{Si}$ located at binding energies of $E_B = 100.07 \pm 0.05$ eV and $E_B = 100.48 \pm 0.07$ eV, respectively. These new components are shifted equally in binding energy relative to $B_{Si}$ and $Z_{Si}$ by 1.4 ± 0.1 eV ($B_{Si} \rightarrow B'_{Si}$) and 1.35 ± 0.12 eV ($Z_{Si} \rightarrow Z'_{Si}$), and are separated equally in binding energy – 0.36 ± 0.1 eV ($Z_{Si} \rightarrow B_{Si}$) and 0.41 eV ± 0.12 eV ($Z'_{Si} \rightarrow B'_{Si}$) to within error. The $RI_R$ of these components suggests that $B'_{Si}$ ($RI_R$ = 0.35) is in the bulk, while $Z'_{Si}$ ($RI_R$ = 1.1) is closer to the surface. This situation is illustrated with the atomic model inset in Fig. 2d, and further supports that components $B_{Si}$ ($Z_{Si}$) and $B'_{Si}$ ($Z'_{Si}$) are simply binding energy-shifted equivalents of each other. The shift to lower binding energy of SiC components is typically observed in the intercalation of graphene on silicon carbide systems and is caused by significant band bending upon intercalant insertion beneath the buffer layer.[7, 46, 49, 58-59, 62-64]

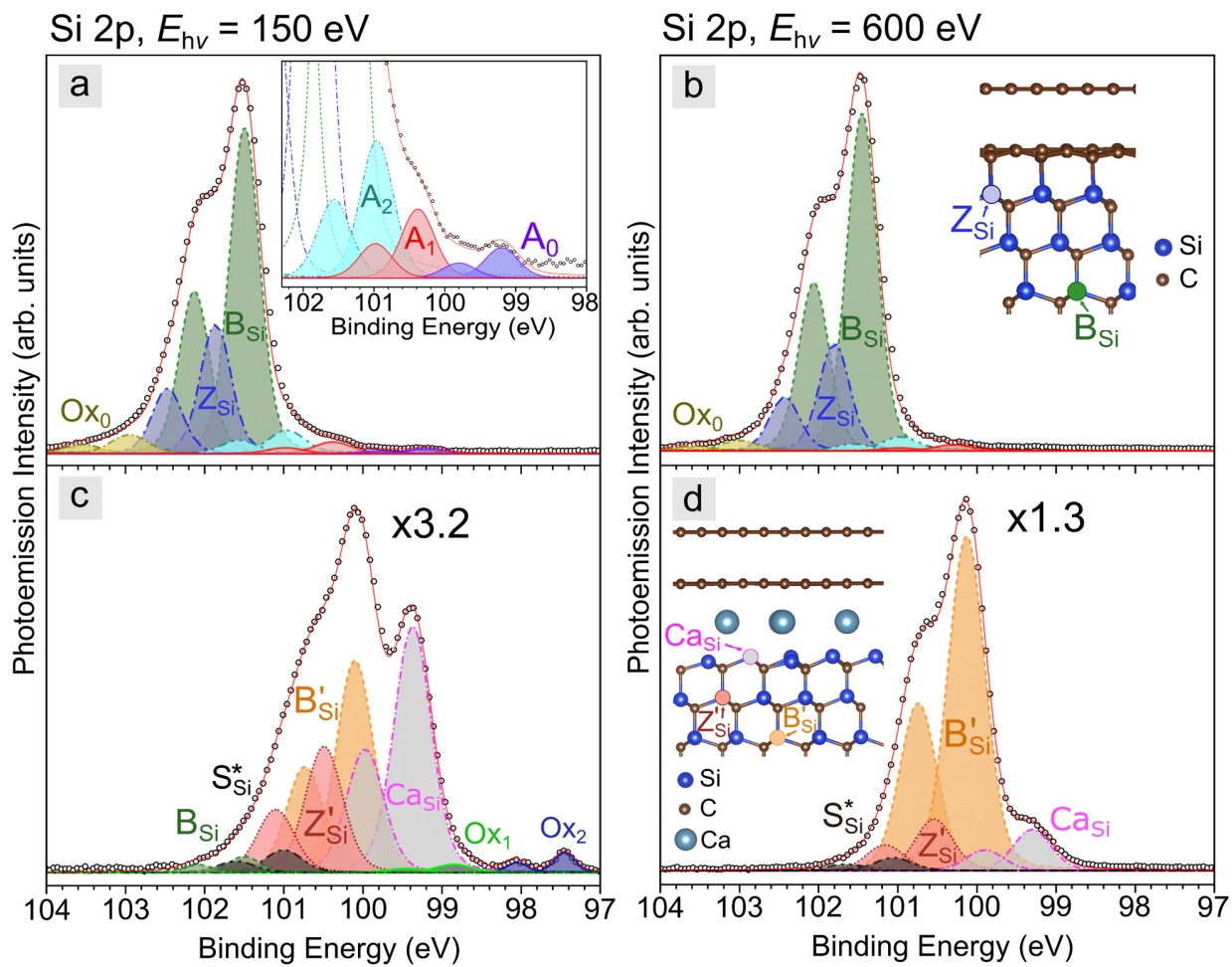

**Figure 2.** X-ray photoemission spectra showing the Si 2p core level of pristine epitaxial monolayer graphene on SiC(0001) (EMLG) and Ca-intercalated EMLG (Ca-QFSBLG) at two incident x-ray energies ($E_{h\nu}$) – 150 eV (left, surface sensitive) and 600 eV (right, bulk substrate sensitive). **(a), (b)** Pristine EMLG before intercalation showing all components. Component $B_{Si}$ is the bulk component of SiC, whereas component $Z_{Si}$ is a more surface SiC component. The approximate atom locations of components $B_{Si}$ and $Z_{Si}$ are shown in the inset of **b**. **(c), (d)** Shows EMLG after Ca-intercalation (Ca-QFSBLG) and new components $Ca_{Si}$, $B'_{Si}$, $Z'_{Si}$ and $S^*_{Si}$. Components $Z_{Si}$ and $B_{Si}$ have been shifted to lower binding energy and are now labelled $Z'_{Si}$ and $B'_{Si}$ respectively. **c** has its y-scale magnified ×3.2 from **a**, and **d** has its y-scale magnified ×1.3 from **b**. The approximate atom positions of the components are shown in the inset of **d**, see text for further explanation of the components.

**Table 1** – Summary of the fitting parameters for the major components in Fig. 2. Both Si 2p$_{3/2}$ and 2p$_{1/2}$ peak locations are given, as well as the relative intensity (*RI*) of the 2p$_{3/2}$ peaks as referenced to the bulk SiC peak 'B$_{Si}$' corresponding to pristine SiC prior to Ca intercalation. $G_W$ refers to the Gaussian FWHM for the Voigt fit (which was fixed to one of two values, except for component Ox$_2$), whereas the Lorentzian FWHM ($W_L$) was kept constant at 0.1 eV in all fits. Dash line corresponds to no observation.

| EMLG, **Si 2p** | | (a), (b), **Pristine EMLG** | | | | (c), (d) **2$^{nd}$ Ca Intercalation** | | | |
|---|---|---|---|---|---|---|---|---|---|
| Component | $E_{hv}$ (eV) | $E_B$ (eV) 2p$_{3/2}$ | $E_B$ (eV) 2p$_{1/2}$ | RI | $G_w$ (eV) | $E_B$ (eV) 2p$_{3/2}$ | $E_B$ (eV) 2p$_{1/2}$ | RI | $G_w$ (eV) |
| B$_{Si}$ | 150 | 101.50 | 102.13 | 1 | 0.4 | 101.45 | 102.05 | 0.016 | 0.4 |
| | 600 | 101.45 | 102.06 | 1 | 0.4 | - | - | - | - |
| Z$_{Si}$ | 150 | 101.86 | 102.47 | 0.40 | 0.4 | | | | |
| | 600 | 101.80 | 102.43 | 0.32 | 0.4 | | | | |
| Ox$_0$ | 150 | 102.95 | 103.55 | 0.06 | 0.5 | | | | |
| | 600 | 102.95 | 103.55 | 0.03 | 0.5 | | | | |
| B$'_{Si}$ | 150 | | | | | 100.11 | 100.74 | 0.24 | 0.5 |
| | 600 | | | | | 100.13 | 100.74 | 0.68 | 0.5 |
| Z$'_{Si}$ | 150 | | | | | 100.51 | 101.13 | 0.12 | 0.5 |
| | 600 | | | | | 100.54 | 101.16 | 0.10 | 0.5 |
| S$^*_{Si}$ | 150 | | | | | 101 | 101.63 | 0.02 | 0.5 |
| | 600 | | | | | 101.05 | 101.65 | 0.026 | 0.5 |
| Ca$_{Si}$ | 150 | | | | | 99.37 | 99.97 | 0.24 | 0.5 |
| | 600 | | | | | 99.3 | 99.9 | 0.08 | 0.5 |
| Ox$_1$ | 150 | | | | | 98.85 | 99.45 | 0.008 | 0.5 |
| | 600 | | | | | - | - | - | - |
| Ox$_2$ | 150 | | | | | 97.45 | 98.05 | 0.02 | 0.25 |
| | 600 | | | | | - | - | - | - |

Figure 3 (component fit parameters shown in Table 2) shows the C 1s core level XPS spectra before and after Ca-intercalation of the same EMLG sample as Fig. 1b (and Fig. 2). An $E_{h\nu}$ of 330 eV and 600 eV was used to characterize surface and bulk components, respectively. Fig 3a, b shows the pristine EMLG sample prior to intercalation, consisting of 5 components. Component G located at $E_B$ = 284.83 ± 0.05 eV, corresponds to graphene, in agreement with prior reports.[65] All graphene-related peaks were fit using a Breit-Wigner-Fano lineshape,[66] instead of the often used Doniach-Šunjić lineshape,[67-68] as the BWF lineshape was found to more accurately describe our data (see Supporting Information, Section 1.1 and 1.3 for further details).

Components S1 ($RI_R$ = 10.48) and S2 ($RI_R$ = 11.73) are due to Si-C bonds from the buffer layer to the underlying SiC and C-C bonds in the buffer layer, respectively,[7, 65] and have an intensity ratio of S2:S1 ≈ 2:1, in agreement with previous findings.[65] Additionally, two components at lower binding energy are resolved – components $Z_C$ and $B_C$. These components are related to the carbon in the SiC surface and bulk respectively. Component $B_C$ at $E_B$ = 283.73 ± 0.05 eV ($RI_R$ = 1, used as the $RI$ reference) represents the well-known EMLG bulk component of SiC, and agrees well in binding energy with prior measurements.[47, 65] We also find evidence for a surface-like component, $Z_C$, related to the carbon in SiC at $E_B$ = 284.05 ± 0.05 eV ($RI_R$ = 1.74) – particularly evident from the asymmetry of the SiC peak in Fig. 3b. Component $Z_C$ has been fitted in previous reports in order to explain the different bonding environment of carbon towards the surface of the SiC.[43, 49] The shift of $Z_C$ relative to $B_C$ of $\Delta E_B$ = 0.32 ± 0.1 eV, is in agreement with the shift observed in the Si 2p spectra ($Z_{Si}$ to $B_{Si}$) and suggests a similar chemical environment for these components. The inset in Fig. 3a shows components $Z_C$ and $B_C$ more clearly at $E_{h\nu}$ = 150 eV, and the inset of Fig. 3b shows the approximate location of all components in an atomic model.

Figures 3c and 3d show the C 1s core level XPS spectra for the same sample after Ca-intercalation. It is immediately apparent that components S1 and S2 are no longer present, implying the absence (or near elimination of) bonding between the SiC and buffer layer. Moreover, although significantly suppressed in intensity, components $B_C$ and $Z_C$ are still present in the Ca-intercalated spectra. This is in agreement with the Si 2p core level spectra in Fig. 2c, d (which showed the presence of component $B_{Si}$) and LEED (Fig. 1b) suggesting partial intercalation, similar to that observed with hydrogen intercalation.[41]

The graphene peak, now labelled G' located at $E_B$ = 284.88 ± 0.05 eV ($RI_R \approx 13$), is not shifted significantly (within measurement error) with respect to the original EMLG graphene component G ($E_B$ = 284.83 ± 0.05 eV), although from prior Ca-intercalation experiments,[8,12] we expect significant n-type doping of the graphene. We do on the other hand observe significant changes in the lineshape of the graphene component, which is significantly broadened and more asymmetric. This increase in asymmetry of the G' peak is predicted to arise in highly doped graphene due to 2D plasmon losses[69-70] (see Supporting Information Section 1.2/1.4 and Methods for details on G' peak fitting), and suggests that the (n-type, as we shall soon discover) doping level of the graphene has increased significantly.

Nonetheless, we observe that the bulk component $B_C$ and the associated surface component of the bulk, $Z_C$, *have* been shifted to lower binding energy – labeled as components $B_C^{'}$ ($E_B$ = 282.33 ± 0.05 eV, $RI_R \approx 0.72$) and $Z_C^{'}$ ($E_B$ = 282.70 ± 0.05 eV, $RI_R \approx 1.34$). The respective shifts in $E_B$ are -1.40 ± 0.1 eV ($B_C \rightarrow B_C^{'}$), -1.35 ± 0.1 eV ($Z_C \rightarrow Z_C^{'}$), with a separation between components of 0.32 ± 0.1 eV ($Z_C \rightarrow B_C$) and 0.37 ± 0.1 eV ($Z_C^{'} \rightarrow B_C^{'}$). These shifts are in good agreement with the shifts in $E_B$ for the analogous Si 2p components in the preceding discussion, and thus gives confidence in our assignment of component $B'_C/B'_{Si}$ and $Z'_C/Z'_{Si}$ as the shifted components $B_C/B_{Si}$ and $Z_C/Z_{Si}$ brought about by band bending effects induced by the creation of a Ca-Si layer on the SiC(0001) surface. Approximate atom locations are shown in the atomic model inset of Fig. 3d.

An additional component, which we label $S^*_C$, is evident in the Ca-intercalated spectra, located at $E_B = 283.26 \pm 0.05$ eV, shifted from component $B_C$ by $0.47 \pm 0.1$ eV. An analogous component was observed in the Si 2p core level spectrum (Fig. 2c, d), $S^*_{Si}$ at $E_B = 101.01 \pm 0.05$ eV, shifted to higher binding energy from component $B_{Si}$ by $0.47 \pm 0.1$ eV. This equivalence in binding energy shift across both core levels suggests that these components are from the same SiC compound. The $E_B$ value of component $S^*_C$ agrees well with sub-stoichiometric values of SiC.[71] Indeed, judging from Fig. 3c, d, we see that $S^*_C$ ($RI_R = 8.25$) is a surface component (hence the "S" assignment), and thus could be a result of SiC unreacted with Ca, yielding a C-rich SiC compound residing at the surface (and thus, partly explaining the relatively smaller, $RI_R \approx 1$, in the Si 2p core level). Further discussion of component $S^*_C$ can be found in Supporting Information, Section 1.2.

In summary, the Si 2p and C 1s core level spectra have yielded strong evidence for: (1) a Ca-Si interaction at the surface of the SiC(0001), and (2) the elimination of buffer layer-SiC bonding. The formation of a Ca-Si compound is also supported by the Ca 2p spectra which can be found in the Supporting Information, Section 1.5. Coupled with the LEED results of Fig. 1b, we conclude that Ca intercalates between the buffer layer and SiC(0001), bonding with the silicon at the SiC(0001) surface, disrupting the buffer layer-SiC bonding and converting the buffer layer into a new graphene layer.

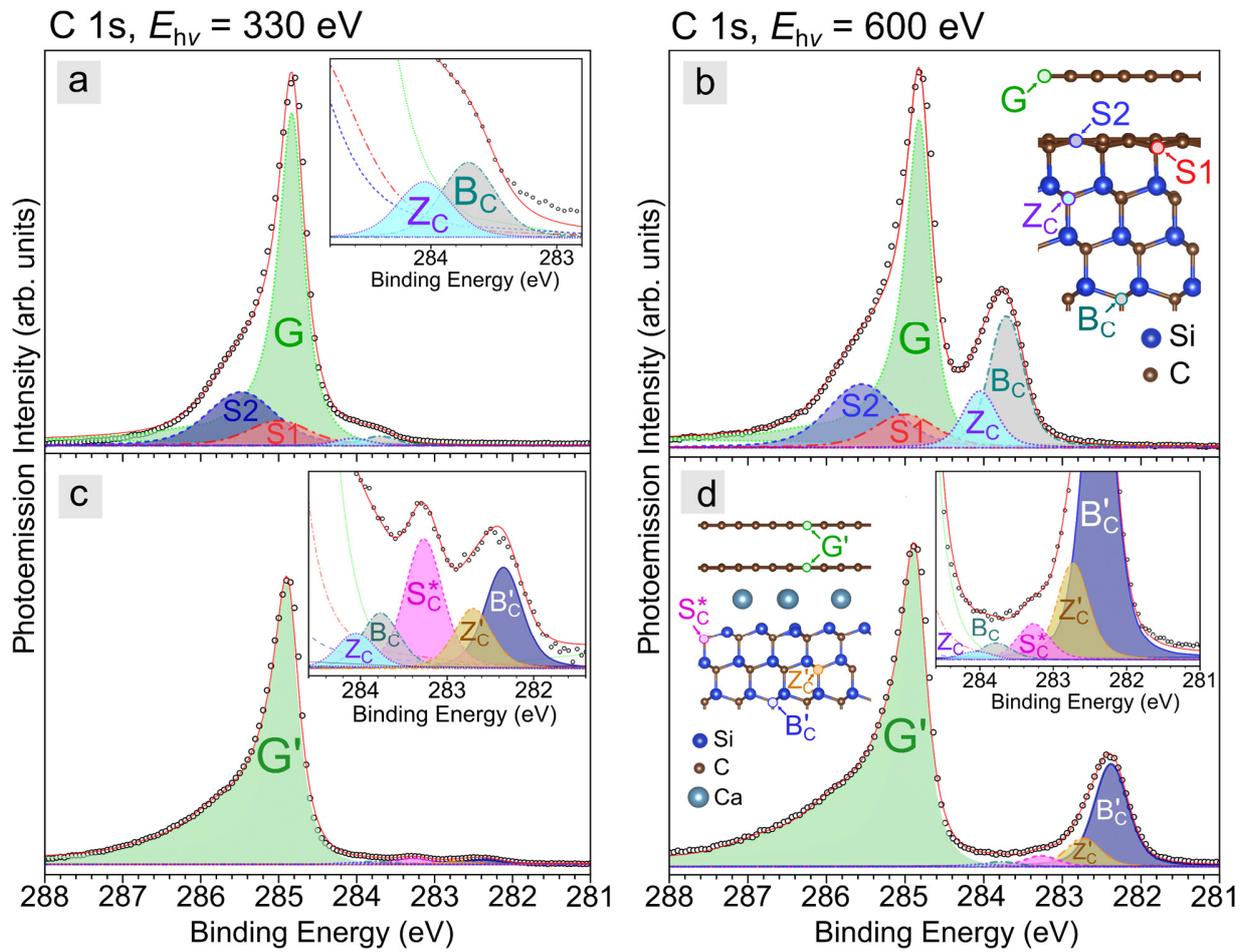

**Figure 3.** X-ray photoemission spectra showing the C 1s core-level of pristine epitaxial monolayer graphene on SiC(0001) (EMLG) and Ca-intercalated EMLG (Ca-QFSBLG) at two x-ray energies, $E_{h\nu} = 330$ eV (left, surface sensitive) and $E_{h\nu} = 600$ eV (right, bulk substrate sensitive). **(a), (b)** Pristine EMLG before Ca-intercalation showing deconvoluted components G (graphene), S1/S2 (carbon in the buffer layer) and $Z_C/B_C$ (surface/bulk SiC). The SiC components at lower binding energy are more clearly seen in the magnified view of the inset of Fig. 3a. The approximate atomic location of these components is shown in the inset of **b** (see text for details). **(c), (d)** Ca-QFSBLG showing component G' (doped graphene). Three additional SiC components have emerged – $S_C^*$, $Z_C^{'}$ and $B_C^{'}$, related to the changed SiC conditions due to formation of a Ca-Si compound between the SiC and buffer layer. The SiC components at lower binding energy are magnified in the right insets. The approximate atom locations for the new components are shown in the left inset of **d**.

**Table 2** – Summary of the fitting parameters for the major components in Fig. 3. The binding energy location ($E_B$) is stated, as well as the relative intensity (*RI*) of the 1s peaks as referenced to the bulk SiC peak '$B_C$' corresponding to pristine SiC prior to Ca-intercalation. '*Q*' refers to the asymmetry parameter of the graphene, which was fit using a Breit-Wigner-Fano (BWF) function. '$W_G$' refers to the gaussian FWHM for the Voigt fit, whereas the Lorentzian FWHM ($W_L$) was kept constant at 0.2 eV in all fits except for the buffer layer components S1 and S2.

| EMLG, C 1s | | (a), (b), **Pristine EMLG** | | | (c), (d) **2ⁿᵈ Ca Intercalation** | | |
|---|---|---|---|---|---|---|---|
| Component | $E_{h\nu}$ (eV) | $E_B$ (eV) | *RI* | $Q/W_L$ or $W_G/W_L$ (eV) | $E_B$ (eV) | *RI* | $Q/W_L$ or $W_G/W_L$ (eV) |
| S2 | 330 | 285.48 | 5.63 | 0.8/0.4 | | | |
| | 600 | 285.55 | 0.48 | 0.8/0.4 | | | |
| S1 | 330 | 285.00 | 2.62 | 0.7/0.4 | | | |
| | 600 | 285.00 | 0.25 | 0.7/0.4 | | | |
| G | 330 | 284.83 | 35.0 | 12/0.2 | | | |
| | 600 | 284.82 | 2.47 | 9/0.2 | | | |
| G' | 330 | | | | 284.89 | 26.75 | 9/0.2 |
| | 600 | | | | 284.88 | 2.09 | 9/0.2 |
| $Z_C$ | 330 | 284.05 | 0.75 | 0.4/0.2 | 284.05 | 0.18 | 0.4/0.2 |
| | 600 | 284.05 | 0.43 | 0.4/0.2 | 284.05 | 0.02 | 0.4/0.2 |
| $Z_C'$ | 330 | | | | 282.70 | 0.31 | 0.4/0.2 |
| | 600 | | | | 282.73 | 0.22 | 0.4/0.2 |
| $B_C$ | 330 | 283.7 | 1 | 0.4/0.2 | 283.77 | 0.29 | 0.4/0.2 |
| | 600 | 283.71 | 1 | 0.4/0.2 | 283.78 | 0.04 | 0.4/0.2 |
| $B_C'$ | 330 | | | | 282.35 | 0.52 | 0.4/0.2 |
| | 600 | | | | 282.38 | 0.77 | 0.4/0.2 |
| $S_C^*$ | 330 | | | | 283.27 | 0.66 | 0.4/0.2 |
| | 600 | | | | 283.27 | 0.08 | 0.4/0.2 |

**Low Temperature Scanning Tunneling Microscopy of Ca-QFSBLG.** Figure 4a shows an STM image of pristine EMLG prior to Ca-intercalation and transformation to Ca-QFSBLG. Apparent is the atomic resolution and Moiré pattern of the underlying buffer layer showing the (6√3×6√3)R30° and associated (6×6) periodicity as expected.[39, 72-73] The inset in Fig. 4a shows the fast Fourier transform (FFT) of the STM image. The graphene spots, labeled G(1×1), yield a lattice parameter of 0.253 ± 0.005 nm, in reasonable agreement with the lattice parameter of graphene (0.246 nm). In addition, we observe the (6√3×6√3)R30° spots around the G(1×1) spots, in agreement with the pristine LEED images in Fig. 1a. Another feature in Fig. 4a are the bright spots (circled in Fig. 4a) scattered randomly on the surface of the EMLG. These bright spots have been previously observed, and have been attributed to silicon adatoms on the surface of the SiC.[74] This is further strengthened by our interpretation that the spectral components $A_0$, $A_1$ and $A_2$, observed in the Si 2p photoemission spectra of pristine EMLG (Fig. 2a, b), as corresponding to Si adatoms (See Supporting Information Section 1.2). Further STM data on the bright spots, including atomic resolution and height measurement, is presented in Supporting Information Section 1.6.

Figure 4b shows the same sample (from Fig. 4a) after partial Ca intercalation as imaged with the STM at 0.12 V, 300 pA (see Supporting Information Section 1.6 for further details on Ca deposition). What is immediately apparent is that the surface is now covered by a network of raised 'islanded' areas which no longer show the long-wavelength (6√3×6√3)R30° modulation associated with the buffer layer. This observation is consistent with previous alkali[52-53] and hydrogen[75] intercalation studies of graphene on SiC, and indicates the formation of a freestanding structure. This observation is also consistent with the LEED data in Figure 1 and with the XPS data in Figures 2 and 3 which imply that the Si-C bonds that define the buffer layer are largely eliminated *via* the suppression of the (6√3×6√3)R30°/(6×6) spots and formation of a Ca-Si compound (component Ca$_{Si}$)/elimination of components S1 and S2, respectively. The height of the newly formed feature is ≈ 0.26 ± 0.01 nm as judged by a histogram of heights from the STM image shown in the inset of Figure 4b. This is in agreement with the displacement measured by Fiori *et al.*[53] after the intercalation of Li atoms underneath the buffer layer, and also agrees with the theoretical height increase for Ca-QFSBLG (see below).

Figure 4c shows an STM image of the same sample after another intercalation step, imaged at 0.1 V and 400 pA. Imaging the sample became increasingly challenging after each intercalation step due to the difficulty of removing excess Ca, which formed clusters on the graphene surface. A method for the removal of excess Ca on the surface using the STM tip was developed, and resulted in atomically clean surfaces (see Supporting Information Section 1.6 for details). Nonetheless, Fig. 4c shows that after the second intercalation, entire terraces now lack the (6√3×6√3)R30° modulation; we again interpret these as Ca-intercalated (*i.e.* Ca-QFSBLG) regions.

Figure 4d shows the FFT of the surface in Fig. 4c and elucidates the underlying symmetry of the surface. Here we can see graphene spots (G(1×1)), SiC spots (SiC(1×1)) and the (6√3×6√3)R30° symmetry that still exists, in agreement with the LEED from Fig. 1. Notably however, unlike the LEED results which showed (√3×√3)R30° spots with respect to the G(1×1) spots, Fig. 4d shows no evidence of the G(√3×√3)R30° symmetry. This provides strong evidence that the graphene bilayer created by Ca-intercalation is not itself intercalated with Ca to form CaC$_6$/C$_6$CaC$_6$,[19] which has a unit cell that is (√3×√3)R30° with respect to the graphene. Thus, we conclude that Ca intercalation has formed quasi-freestanding bilayer graphene (*i.e.* Ca-QFSBLG), separated from the SiC surface by a Ca layer which has interacted strongly with the Si on the SiC(0001) surface. The ($\sqrt{3}\times\sqrt{3}$)R30° spots observed in LEED indicate that the Ca underneath the QFSBLG is indeed in registry with the graphene, however the STM measurements indicate that the perturbation of the ($\sqrt{3}\times\sqrt{3}$)R30° Ca layer on the bilayer graphene unit is small.

The Ca-intercalated graphene areas in Figs. 4b and 4c are different in appearance to the pristine EMLG starting sample in Fig. 4a and contain 'dark' *and* 'bright' spots (different to the 'bright spots attributed to Si adatoms prior to Ca-intercalation) on the surface (these spots are particularly evident on the uniform terrace in Fig. 4c, but can also be seen in the partially intercalated 'island' areas in Fig. 4b). The dark spots (which are not holes – see Supporting Information Section 1.6) have a height variation of <40 pm (see Fig. 4c inset), and thus are likely not due to the buffer layer corrugation which is ≈ 60 pm[73] – in agreement with the height variation of similar features measured with STM after H-intercalation of graphene on SiC.[41] In fact, similar 'dark' features have been observed using STM after alkali,[52-53] indium[43] and hydrogen[41] intercalation of graphene on silicon carbide. The observation of these features is thus, intercalant *independent*, and points to a universal underlying mechanism involving the SiC surface. In the case of H-intercalation, Murata *et al.*[41] attributed these features to hydrogen vacancies, with a measured density ranging from ≈ $10^{11}$ to $10^{13}$ cm$^{-2}$ depending on the intercalation temperature used. Here we find that our dark spot density of ≈ $9\times10^{12}$ cm$^{-2}$ is within this range, and agrees favorably to the dark spot density found after hydrogen intercalation.[41] The density of silicon adatoms (bright spots in Fig. 4a) is on the order of $10^{13}$ cm$^{-2}$ for our measured data, whereas Rutter *et al.*[74] observed Si adatom densities on the order of $10^{14}$ cm$^{-2}$, in rough agreement with our numbers. Thus, we propose that the dark spots are Ca vacancies that are tied to Si adatom locations on the surface of the SiC. As for the post-intercalation bright spots (arrow in Fig. 4c), the height variation does not exceed ≈ 30 pm (see Supporting Information Section 1.6) and the density is ≈ $1\times10^{12}$ cm$^{-2}$, in approximate agreement with prior measurements on H-intercalated graphene on SiC.[41] We note here that the post-intercalation bright spots we observe after Ca-intercalation are likely not the same bright spots that were initially attributed to silicon adatom defects in the STM micrograph of pristine EMLG in Fig. 4a. Previous reports have attributed these bright spots to defects[41] or protrusions from intercalants bonded to the underlying graphene, rather than the SiC surface.[76] Since we do not observe electron scattering standing waves from the post-intercalation bright spots,[77-78] our observations suggest the latter. Thus, we tentatively attribute these bright spots to excess metallic Ca that may be present on top of the underlying Ca-silicide. For further information on these post-intercalation bright spot features, we refer the reader to Supporting Information Section 1.6. We mention here briefly that the STM of Ca-intercalated graphene has been previously conducted,[19, 21, 25] but the bright and dark spot features shown in Figures 4b and c were not (to our best knowledge) reported.

The STM has elucidated the surface structure of Ca-intercalated EMLG on SiC(0001). In summation, we have found that Ca-QFSBLG has formed upon intercalation of Ca via elimination of EMLG's Moiré pattern, and, that no Ca-intercalation between the graphene layers occurred – consistent with our LEED (Fig. 1b) and XPS (Figs. 2 and 3) results.

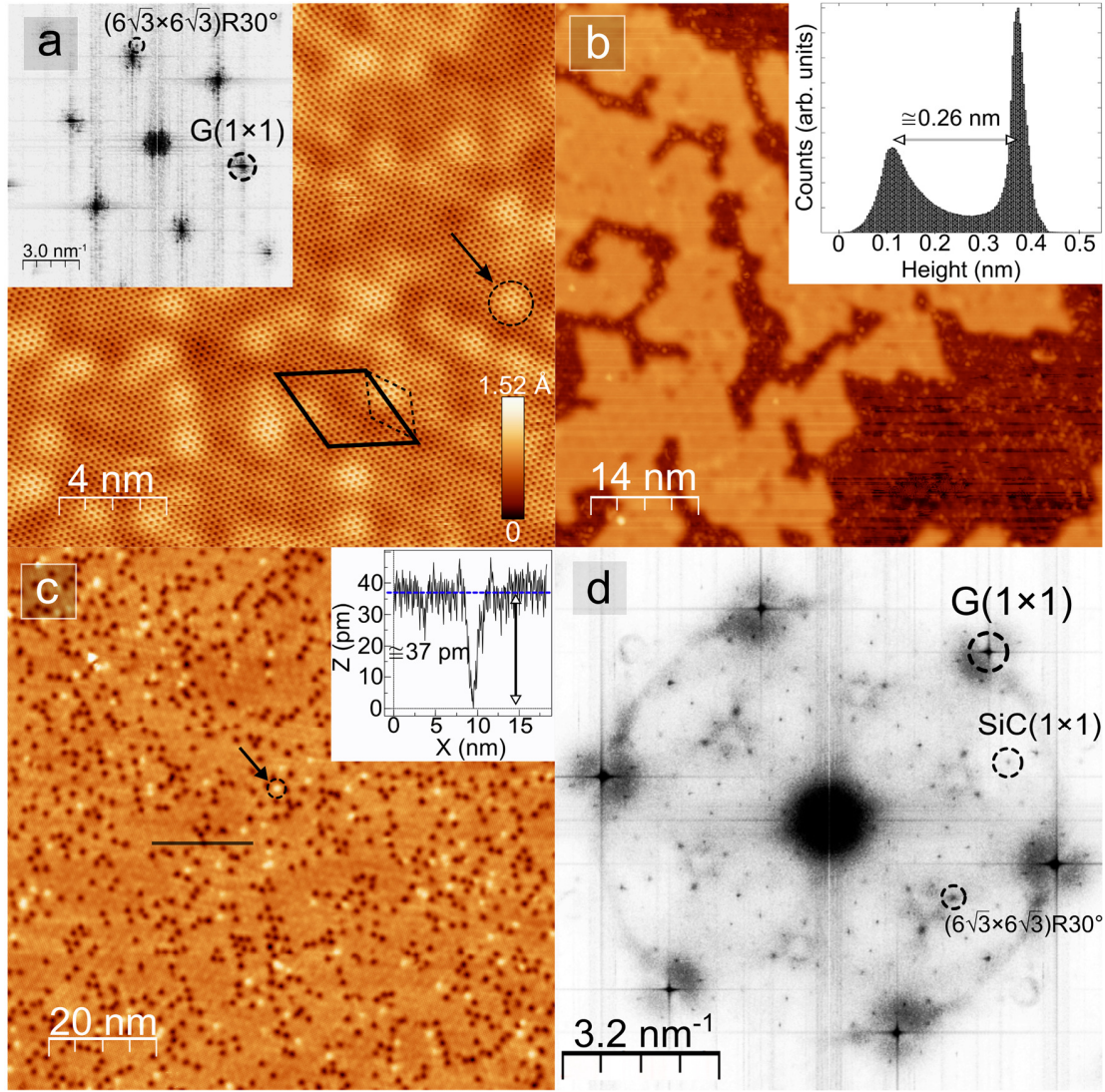

**Figure 4.** Scanning tunneling microscopy (STM) images of pristine epitaxial monolayer graphene (EMLG) before and after Ca-intercalation. **(a)** Pristine EMLG (1 V, 200 pA, 77 K) showing atomic resolution, along with real space (6√3×6√3)R30° (solid line) and (6×6) (dashed line) quasi-periodicities (Moiré). The bright spots correspond to silicon adatom features (dotted circle, arrow). Inset shows the fast fourier transform (FFT) of same image showing G(1×1) graphene spots with measured lattice parameter of ≈ 0.25 nm, and (6√3×6√3)R30° spots around G(1×1) spots. **(b)** 1$^{st}$ Ca-intercalation step (0.12 V, 300 pA, 4.6 K) showing formation rasied areas (Ca-QFSBLG). The height difference between these raised areas and the un-intercalated graphene is 0.26 ± 0.01 nm (see inset). **(c)** After the 2$^{nd}$ Ca-intercalation, Ca-QFSBLG formed on entire terraces (0.1 V, 400 pA, 4.6 K). Arrow and dotted circle shows post-intercalation bright spot feature (different from bright spot in **a** – see text). Inset shows a topographic line scan across one of the dark spots in the image, denoted by the solid black line. (**d**) FFT of **c** showing underlying symmetry of Ca-QFSBLG surface and graphene (G(1×1)), SiC (SiC(1×1)) and (6√3×6√3)R30° spots.

**Density Functional Theory of Ca-intercalated graphene on SiC(0001).** Our results strongly imply that Ca-intercalation of EMLG occurs between the buffer layer and SiC(0001) surface to form Ca-QFSBLG, and not between the buffer layer and graphene (or between graphene layers). To better understand this, we utilize DFT to investigate the energetics of Ca-intercalation.

Figure 5 shows schematics of the two structures modelled using DFT. Figs. 5a and b show side and top views of the first modelled structure, corresponding to Ca-QFSBLG, *i.e.*, Ca intercalated between the buffer layer – SiC interface to produce bilayer graphene from EMLG (SiC/Ca/Graphene/Graphene). The former buffer layer is raised to lie 0.467 nm above the surface Si atoms, a change in height ($\Delta h$), compared to pristine EMLG, of approximately 0.26 nm (using the known value for buffer layer – Si distance of 0.21 nm),[79] in excellent agreement with our experimentally measured value of $\Delta h = 0.26 \pm 0.01$ nm from STM. Figs. 5c and d show a second modelled structure, corresponding to Ca instead intercalated between the 1st graphene layer and buffer layer (SiC/Buffer Layer/Ca/Graphene), a model that has been, up until now, thought to describe Ca-intercalation.[20] Using this model, the insertion of Ca gives a $\Delta h$ of 0.153 nm when compared with pristine EMLG, which is not in agreement with our STM measurements (see Fig. 4b inset). The energy, $E_I$, required to intercalate Ca was calculated (see Methods) yielding $E_I$ = -2.5 and +1.57 eV/atom for Ca intercalated underneath the buffer layer and Ca intercalated between the 1st graphene layer and buffer layer, respectively. Here, a negative value indicates the process is energetically favorable. Thus, DFT supports our experimental finding that the favored intercalation site for the Ca is between the SiC and the buffer layer.

It should be noted that our DFT results closely match the conclusions of Sandin *et al.*[52] (which used Na as the intercalant), but disagrees with a recent publication by Zhang *et al.*[21] concerning Ca-intercalation. Sandin *et al.*[52] concluded that a (SiC/Na/Graphene/Graphene) structure accurately describes Na-intercalation by comparing experimentally measured and theoretically calculated workfunction shifts. Zhang *et al.*[21] used STM and DFT calculations to find that low concentrations of Ca could intercalate in a 'meta-stable' configuration underneath the buffer layer, although (SiC/buffer layer/Ca/Graphene) and (SiC/buffer layer/Graphene/Ca/Graphene) structures were found to be more energetically favorable – in stark disagreement with our findings which suggest that the (SiC/Ca/Graphene/Graphene) structure is most stable. Furthermore, the change in height after intercalation in reference [21] was experimentally measured as 0.283 nm (theoretically calculated as 0.29 nm), and this disagrees with both of our experimental and theoretical results presented here.

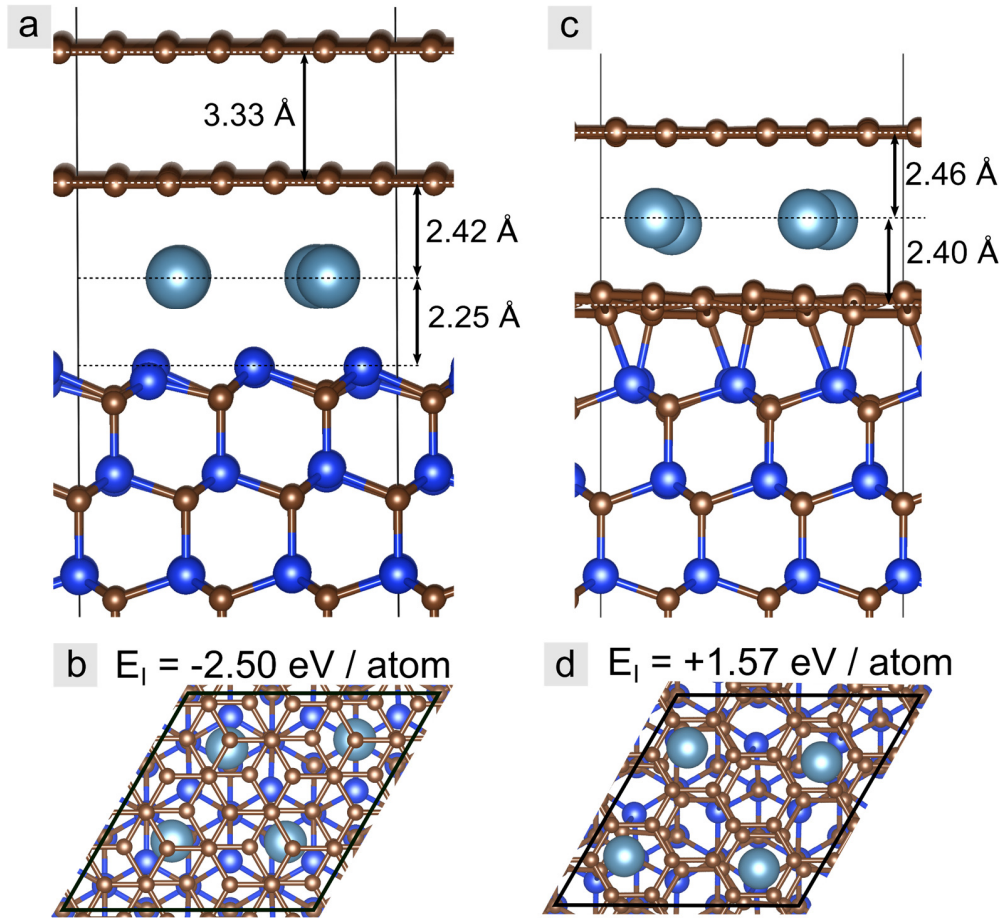

**Figure 5.** Atomic models (dark blue = Si, brown = C, light blue = Ca) of Ca-intercalated monolayer graphene on SiC(0001), showing the two possible intercalation regimes. (**a**) Side view of Ca-QFSBLG showing vertical distances and, (**b**) top down view of Ca-QFSBLG showing the unit cell and approximate ($\sqrt{3}\times\sqrt{3}$)R30° coordination of the Ca with respect to the graphene. The energy required for intercalation, $E_I$ = -2.50 eV / atom (energetically favorable). (**c**) Side view of Ca intercalated between the 1st graphene layer and buffer layer (SiC/Buffer/Ca/Graphene) showing distances and, (**d**) top down view of Ca intercalated between the 1st graphene layer and buffer layer showing unit cell. In the case of (SiC/Buffer/Ca/Graphene) the coordination of the Ca was slightly larger than ($\sqrt{3}\times\sqrt{3}$)R30°. The energy required for intercalation, $E_I$ = +1.57 eV / atom (not energetically favorable).

**X-ray photoelectron spectroscopy of Mg-QFSBLG.** As in the Ca-intercalation case in the above sections, here we will focus on the Si 2p and C 1s core levels shown in Figures 6 and 7, respectively (O 1s and Mg 2p core levels can be found in Supporting Information Section 2.4). All relevant parameters for the component fits can be found in Table 3 (Si 2p) and Table 4 (C 1s). We shall note that the XPS results presented here, as in the Ca-intercalation case, are from the same sample analyzed initially with LEED (Fig. 1c). Figures 6a and 6b show the Si 2p core level of pristine EMLG sample prior to Mg-intercalation, and are almost identical to the spectra obtained for the pristine EMLG sample prior to Ca-intercalation in Figs. 2a, b. Again, the $RI$'s (and thus $RI_R$'s) are referenced to the bulk of the SiC, components $B_{Si}/B_C$, and we retain the same nomenclature as in Fig. 2a, b. All component binding energy location and fit details can be found in the Supporting Information, Tables S10-S12, and since these components have not significantly changed, we refer the reader to the Supporting Information and preceding discussion of Fig. 2a, b.

Figure 6 shows the Si 2p core level spectra before (Figs. 6a and b) and after (Figs 6c and d) Mg-intercalation, showing drastic changes upon intercalation. We discuss again only the main components of the spectra, and leave the discussion of non-essential components to the Supporting Information (see Section 2.2). As we shall soon discover, the changes we observe in the XPS spectra amount to the creation of Mg-QFSBLG, and the XPS spectra are similar to the Ca-QFSBLG sample (with the exception that Mg could not displace the hydrogen in H-QFSBLG, see Supporting Information Section 2.3).

For instance, component $Mg_{Si}$ located at $E_B = 99.94 \pm 0.05$ eV ($RI_R \approx 2.9$) matches closely in binding energy to an Mg-silicide compound.[51, 80-81] We rule out this component arising from an oxide (*i.e.,* from the reaction of Mg with any underlying silicon oxide to form a silicate[82]), since silicon oxides reside at higher binding energy. Thus, our XPS results imply that our Mg deposition and annealing procedure has intercalated Mg between the buffer layer – SiC interface, producing an Mg-Si compound closely resembling an Mg-silicide and decoupling the buffer layer to form another graphene layer – similar to the Ca-intercalation case. This result complements the LEED results in Fig. 1c, and is also supported by Raman mapping spectroscopy data (see Supporting Information Section 2.5).

We also see the appearance of component $Mg_{2s}$ at $E_B = 103.94 \pm 0.05$ eV. A similar component has been observed in previous XPS studies of magnesium silicide formation on Si(111) and Si(100), though its origin is controversial.[80, 83] The observation of a higher binding energy component in the Si 2p core level spectrum after $Mg_2Si$ formation on Si(100) was attributed to plasmon losses from Mg 2s electrons.[83] However, in our experiment there is no observable Mg metal on the surface after the 2$^{nd}$ Mg-intercalation (see the Mg 2p core level in Supporting Information, Fig. S23a-b), and thus, component $Mg_{2s}$ cannot be caused by metallic plasmon losses. In the case of $Mg_2Si$ on Si(111), the higher binding energy component in the Si 2p spectrum ($E_B \approx 104$ eV, in agreement with our measurements) was attributed to satellite peaks from Mg 2s emission.[80] Satellite features can also (in addition to plasmons) be caused by polychromatic light and shake-up features. Since our experiment uses monochromatic light, we tentatively attribute this peak to a shake-up feature. Furthermore, component $Mg_{2s}$ disappears into the signal noise level when $E_{h\nu} = 600$ eV, strongly implying that this component resides near to the surface (*i.e.* is associated with the formation of the Mg-silicide). Further work is needed to confirm the origin of this feature, but its association with magnesium silicide formation lends support to our conclusion that Mg has intercalated at the buffer layer-SiC interface to interact with Si.

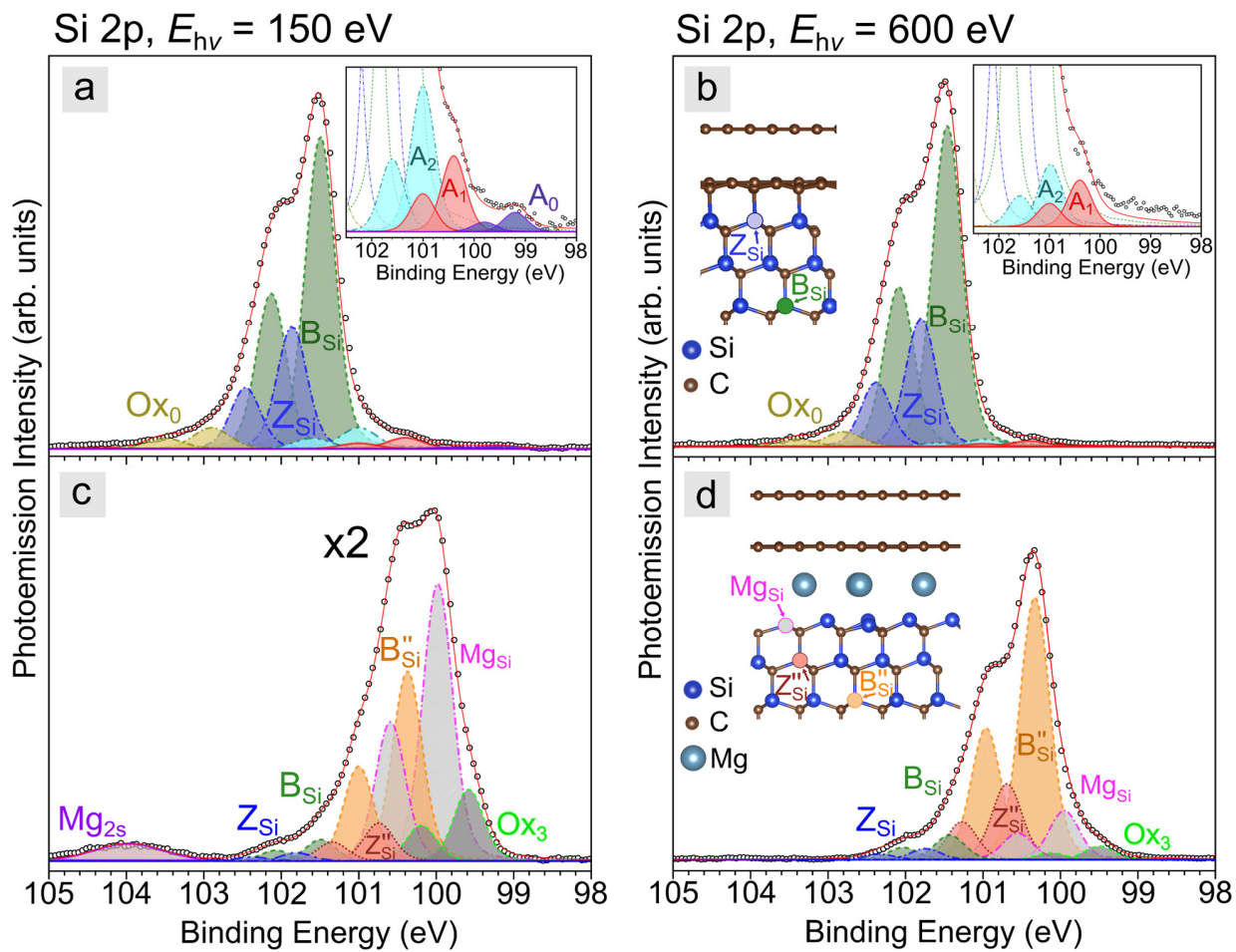

**Figure 6.** X-ray photoemission spectra showing the Si 2p core level of pristine epitaxial monolayer graphene on SiC(0001) (EMLG) and Mg-intercalated EMLG (Mg-QFSBLG) at x-ray energies $E_{h\nu}$ = 150 eV (left, surface sensitive) and $E_{h\nu}$ = 600 eV (right, bulk sensitive). **(a), (b)** Pristine EMLG showing deconvolution of spectra into components (see main text for details). Right hand side inset in **a, b** shows magnified view of lower binding energy components, left hand side inset in **b** shows atomic model and approximate location of EMLG spectral components. **(c), (d)** After Mg-intercalation, Mg-QFSBLG is formed. Spectra are deconvolved into components (see main text for details). Left hand side inset in **d** shows atomic model and approximate location of Mg-QFSBLG spectral components.

**Table 3** – Summary of the fitting parameters for the major components in Fig. 6. Both $2p_{3/2}$ and $2p_{1/2}$ are stated, as well as the relative intensity (*RI*) of the $2p_{3/2}$ peaks as referenced to the bulk SiC peak 'B$_{Si}$' corresponding to pristine SiC prior to Mg intercalation. $W_G$ refers to the gaussian FWHM for the Voigt fit (which was fixed to one of two values), whereas the Lorentzian FWHM ($W_L$) was kept constant at 0.1 eV in all fits. Component Mg$_{2s}$ is not a 2p core level (see text). Dash line corresponds to no observation, n/a = not applicable.

| EMLG, **Si 2p** | | (a), (b), **Pristine EMLG** | | | | (c), (d) **2nd Mg Intercalation** | | | |
|---|---|---|---|---|---|---|---|---|---|
| Component | $E_{hv}$ (eV) | $E_B$ (eV) $2p_{3/2}$ | $E_B$ (eV) $2p_{1/2}$ | *RI* | $W_G$ (eV) | $E_B$ (eV) $2p_{3/2}$ | $E_B$ (eV) $2p_{1/2}$ | *RI* | $W_G$ (eV) |
| Ox$_o$ | 150 | 102.90 | 103.50 | 0.068 | 0.5 | | | | |
|  | 600 | 102.80 | 103.40 | 0.046 | 0.5 | | | | |
| Z$_{Si}$ | 150 | 101.86 | 102.46 | 0.392 | 0.4 | 101.80 | 102.38 | 0.014 | 0.4 |
|  | 600 | 101.80 | 102.38 | 0.401 | 0.4 | 101.74 | 102.34 | 0.035 | 0.4 |
| B$_{Si}$ | 150 | 101.50 | 102.13 | 1 | 0.4 | 101.49 | 102.07 | 0.034 | 0.4 |
|  | 600 | 101.46 | 102.09 | 1 | 0.4 | 101.43 | 102.03 | 0.076 | 0.4 |
| B"$_{Si}$ | 150 | | | | | 100.37 | 101.0 | 0.304 | 0.4 |
|  | 600 | | | | | 100.33 | 100.96 | 0.817 | 0.4 |
| Z"$_{Si}$ | 150 | | | | | 100.74 | 101.34 | 0.061 | 0.4 |
|  | 600 | | | | | 100.69 | 101.27 | 0.235 | 0.4 |
| Mg$_{Si}$ | 150 | | | | | 99.98 | 100.59 | 0.446 | 0.4 |
|  | 600 | | | | | 99.95 | 100.55 | 0.152 | 0.4 |
| Ox$_3$ | 150 | | | | | 99.58 | 100.19 | 0.115 | 0.4 |
|  | 600 | | | | | 99.50 | 100.12 | 0.042 | 0.4 |
| Mg$_{2s}$ | 150 | | | | | 103.97 | n/a | n/a | 1 |
|  | 600 | | | | | - | n/a | n/a | - |

Figures 7a and 7b show the pristine EMLG C 1s core level spectra, and are almost identical in nature to the Ca-intercalation case in Fig. 3a, b. Since the spectrum in Fig 7a, b is almost identical to that of Fig. 3a, b, we refer the reader to the preceding discussion of Fig. 3a, b and Supporting Information Section 1.2/2.2 for further details.

As has been stated, the Mg-intercalation shown in Figs. 6c and 6d, shares similar features to the Ca-intercalation experiment. Upon comparison of Figures 7c and 7d with Figs. 3c and 3d, we observe a similar change in shape and lack of peak shift of the graphene component G, now labelled G'', which we attribute to plasmonic effects due to high doping[69-70] (see Supporting Information Section 1.4) and a reduction of the buffer layer contribution. Additionally, we observe the new components $B''_{Si}/B''_C$ and $Z''_{Si}/Z''_C$ arise after intercalation in the Si 2p and C 1s core level spectra, respectively. These components are analogous to the shifted bulk ($B'_{Si}/B'_C$) and sub-surface ($Z'_{Si}/Z'_C$) SiC components in Ca-QFSBLG (see Figs. 2c, 2d, 3c and 3d), which arise from surface band bending caused by formation of an Mg layer underneath the buffer layer (analogous to Ca as in Figures 2 and 3). The shift of SiC components towards lower binding energy due to band bending at the SiC surface has been observed in many other graphene on SiC intercalation experiments.[7, 46, 49, 58-59, 62-64] The relative shifts in binding energy between components $B_{Si}/B_C$, $Z_{Si}/Z_C$, $B''_{Si}/B''_C$ and $Z''_{Si}/Z''_C$ can be found in Supporting Information Section 2.2, and further supports our assignment of these components.

After analysis of the C 1s and Si 2p XPS spectra, we find remarkable similarities between Ca- and Mg-intercalated graphene. In the case of Mg intercalation, we find (as in the Ca-intercalation case): (1) formation of an Mg-Si compound underneath the buffer layer at the SiC(0001) surface and (2) the disappearance or diminishment of the buffer layer contribution in the Mg-intercalated C 1s spectra. These conclusions are further strengthened by the LEED results in Fig. 1c, and we conclude here that Mg intercalated between the buffer layer and SiC(0001), interacting strongly with the SiC surface and disrupting the buffer layer-SiC bonding to convert the buffer layer into a new graphene layer (Mg-QFSBLG).

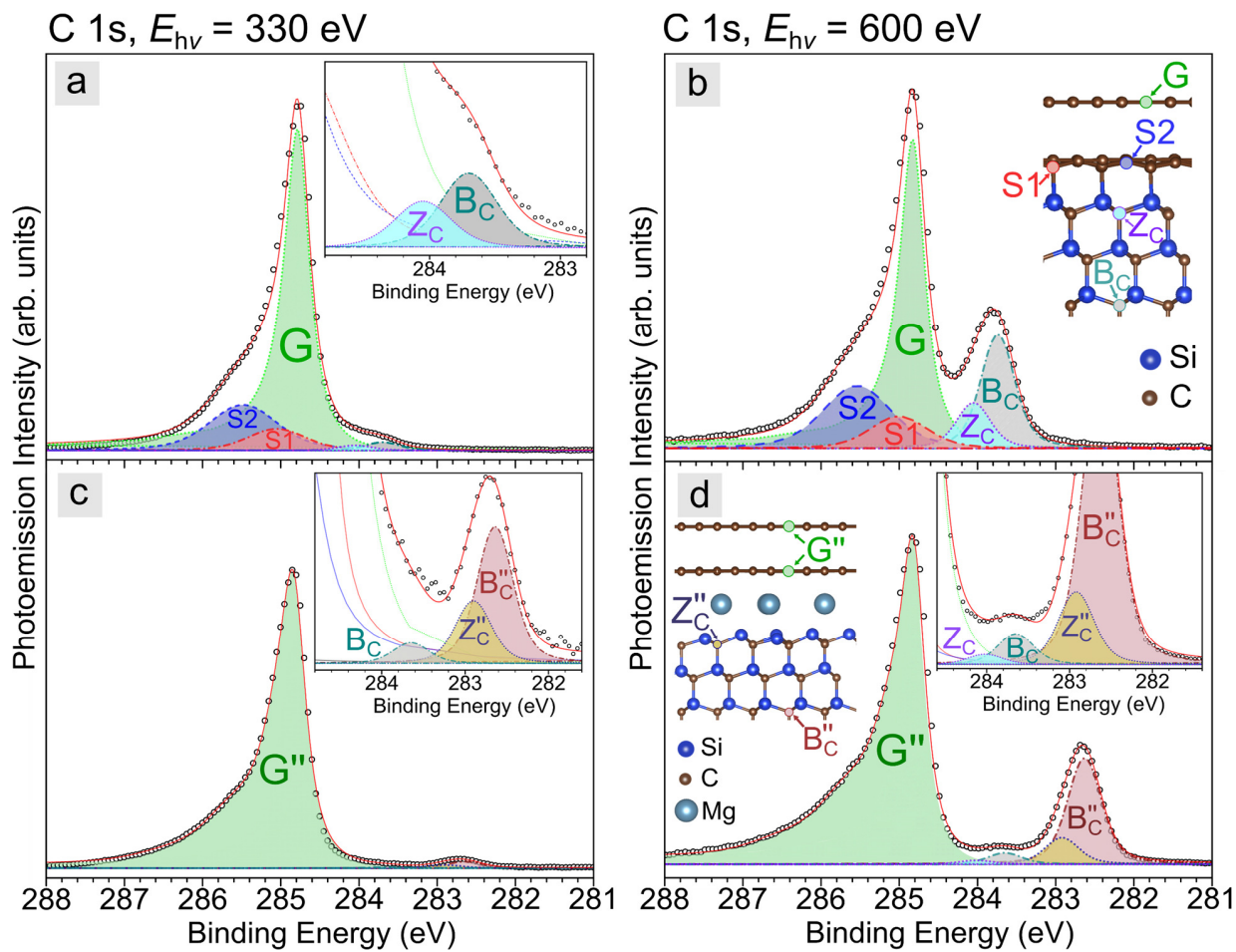

**Figure 7.** X-ray photoemission spectra for the C 1s core level of pristine epitaxial monolayer graphene on SiC(0001) (EMLG) and Mg-intercalated EMLG (Mg-QFSBLG) at x-ray energies of $E_{h\nu}$ = 330 eV (left, surface sensitive) and $E_{h\nu}$ = 600 eV (right, bulk sensitive). **(a), (b)** Pristine EMLG prior to Mg-intercalation showing graphene (G), buffer layer (S1, S2) and SiC ($Z_C$, $B_C$) components. Inset in **a** shows magnified view of $Z_C$ and $B_C$ components. Inset in **b** shows the approximate location of all components with respect to an atomic model of EMLG. **(c), (d)** After Mg-intercalation and the formation of Mg-QFSBLG showing new components, G" (doped graphene), $B_C^{"}$ (shifted bulk SiC component) and $Z_C^{"}$ (shifted 'surface' SiC component). Components $Z_C$ and $B_C$ are still present as the sample was only partially intercalated. Right hand side inset in **c** and **d** shows a magnified view of components $B_C^{"}$ and $Z_C^{"}$. Left hand side inset in **d** shows the approximate location of components $B_C^{"}$ and $Z_C^{"}$.

**Table 4 -** Summary of the fit parameters of the major components in Fig. 3. The binding energy location ($E_B$) is stated, as well as the relative intensity (*RI*) of the 1s peaks as referenced to the bulk SiC peak '$B_C$' corresponding to pristine SiC prior to Ca intercalation. '*Q*' refers to the asymmetry parameter of the graphene (which was fit using a BWF function). '$W_G$' refers to the gaussian FWHM for the Voigt fit, whereas the Lorentzian FWHM ($W_L$) was kept constant at 0.2 eV in all fits except for buffer layer components S1 and S2. Dash line corresponds to no observation.

| EMLG, C 1s | | (a), (b), **Pristine EMLG** | | | (c), (d) **2nd Mg Intercalation** | | |
|---|---|---|---|---|---|---|---|
| Component | $E_{hv}$ (eV) | $E_B$ (eV) | RI | $Q/W_L$ or $W_G/W_L$ (eV) | $E_B$ | RI | $Q/W_L$ or $W_G/W_L$ (eV) |
| S2 | 330 | 285.48 | 5.563 | 0.8/0.4 | | | |
| | 600 | 285.54 | 0.550 | 0.8/0.4 | | | |
| S1 | 330 | 285.05 | 2.636 | 0.7/0.3 | | | |
| | 600 | 285.005 | 0.281 | 0.7/0.3 | | | |
| G | 330 | 284.787 | 39.31 | 12/0.2 | | | |
| | 600 | 284.830 | 2.723 | 12/0.2 | | | |
| G" | 330 | | | | 284.836 | 36.66 | 12/0.2 |
| | 600 | | | | 284.822 | 2.898 | 9/0.2 |
| $Z_C$ | 330 | 284.05 | 0.625 | 0.4/0.2 | - | - | - |
| | 600 | 284.05 | 0.400 | 0.4/0.2 | 284.02 | 0.033 | 0.4/0.2 |
| $Z_C"$ | 330 | | | | 282.90 | 0.375 | 0.4/0.2 |
| | 600 | | | | 282.92 | 0.233 | 0.4/0.2 |
| $B_C$ | 330 | 283.70 | 1 | 0.4/0.2 | 283.65 | 0.125 | 0.4/0.2 |
| | 600 | 283.74 | 1 | 0.4/0.2 | 283.65 | 0.100 | 0.4/0.2 |
| $B_C"$ | 330 | | | | 282.64 | 0.900 | 0.4/0.2 |
| | 600 | | | | 282.63 | 0.933 | 0.4/0.2 |

**Workfunction measurements of Ca- and Mg-QFSBLG.** As noted above, we did not observe significant shifts in the binding energy of the graphene C 1s components upon either Ca- or Mg-intercalation. However, the shift in binding energy of the graphene component is not directly proportional to the shift in the Fermi level ($E_F$).[70, 84] The significant asymmetric broadening of the graphene C 1s component does suggest significant doping of the graphene, which would be expected upon intercalation of an electron donor such as Ca or Mg.[8] A comparison between our experimentally measured C 1s data, and the calculated lineshape of highly doped graphene can be found in Supporting Information Section 1.4, and shows that the lineshape of our data closely matches bilayer graphene doped with at least one layer at $10^{14}$ carriers cm$^{-2}$. To gain further experimental insight into the changes in doping upon intercalation, the workfunction ($W_f$) was measured using the secondary electron cut-off (SECO) photoemission technique, shown in Figure 8.

Figure 8a shows the $W_f$ of pristine EMLG and Ca-intercalated graphene, *i.e.* Ca-QFSBLG (a -9V potential was applied to the samples, see Methods). Our measured value for the $W_f$ of pristine EMLG ($W_f$ = 3.96 ± 0.05 eV) is in approximate agreement with prior values.[85] After Ca-intercalation to form Ca-QFSBLG, the change in workfunction is $\Delta W_f$ = -0.28 ± 0.1 eV and strongly implies an increase in n-type doping.[86] In the case of Mg-intercalation to form Mg-QFSBLG, as shown in Fig. 8b, the change in workfunction from EMLG is $\Delta W_f$ = -0.18 ± 0.1 eV; also implying a significant increase in n-type doping. Further discussion pertaining to the accuracy of the SECO measurements can be found in Supporting Information Section 1.2. The results for Ca-intercalated H-QFSBLG (to form Ca-QFSBLG) are shown in the Supporting Information, Section 1.3, and are almost identical to the results obtained for Ca-intercalated EMLG shown here.

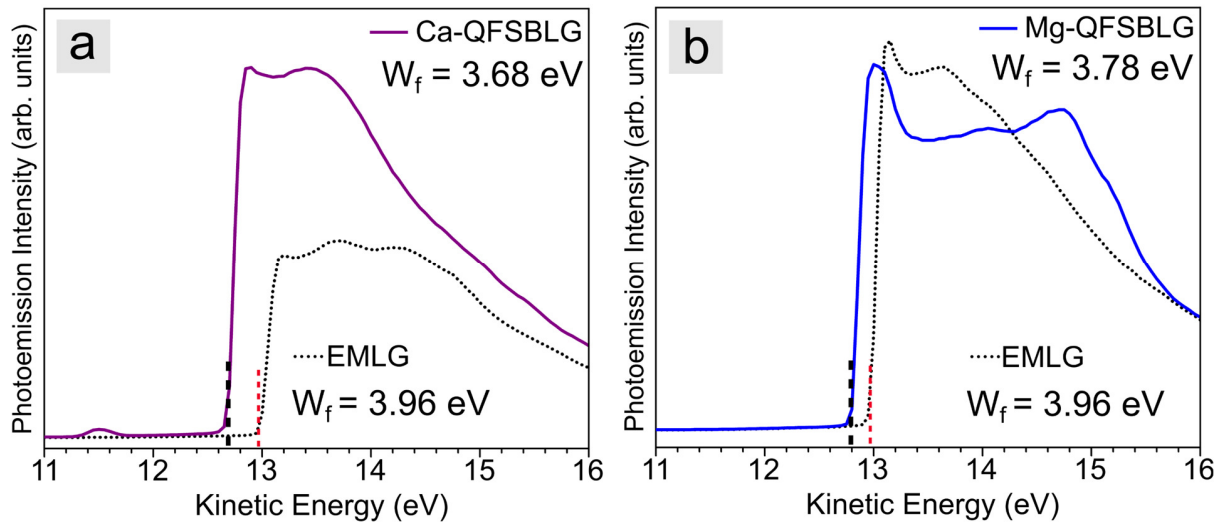

**Figure 8.** Secondary electron cut-off (SECO) photoemission measurements of pristine epitaxial monolayer graphene on SiC (EMLG) before and after Ca- and Mg-intercalation to form quasi-freestanding bilayer graphene on SiC (Ca-QFSBLG/Mg-QFSBLG). **(a)** Pristine EMLG (dotted line) and Ca-QFSBLG (solid purple line) showing a workfunction ($W_f$) of 3.96 and 3.68 eV, respectively. **(b)** Clean EMLG (dotted line) and Mg-QFSBLG (solid blue line) showing a $W_f$ of 3.96 and 3.78 eV, respectively.

**Air stability of Mg-QFSBLG.** All the experiments presented thus far have been conducted in UHV, and with few exceptions,[87] highly doped intercalated graphene is not typically stable in ambient conditions due to the high reactivity of the intercalants used. Whether the applications of such intercalated materials are superconductors, batteries, or highly conductive surfaces – air-stability is an important consideration, significantly aiding in the technological implementation of these materials. As expected, we found our Ca-QFSBLG samples begin oxidation within 30 minutes of ambient exposure, manifesting in a radical change of the XPS spectra. Surprisingly, we find that our newly synthesized Mg-QFSBLG sample appears to be much more stable to air exposure.

Figure 9 shows the Si 2p and C 1s core level spectra for the same sample shown in Figs. 6 and 7 after exposure to ambient atmosphere for ≈ 6 hours, followed by re-introduction to vacuum and subsequent annealing to remove surface contaminants (see Methods). We do not observe a significant shift in any of the Si 2p or C 1s components for our Mg-intercalated sample (see Supporting Information Tables S13-14), suggesting that Mg-intercalated graphene is remarkably stable in ambient conditions. Upon closer inspection, we observe that component $Mg_{Si}$ has decreased in intensity with respect to the other Si 2p components, as compared to the spectrum before ambient exposure (Fig. 6c) suggesting that there may have been some reaction with ambient. Nonetheless, it appears that Mg-intercalation yields a very low workfunction (3.78 eV) yet relatively air-stable graphene.

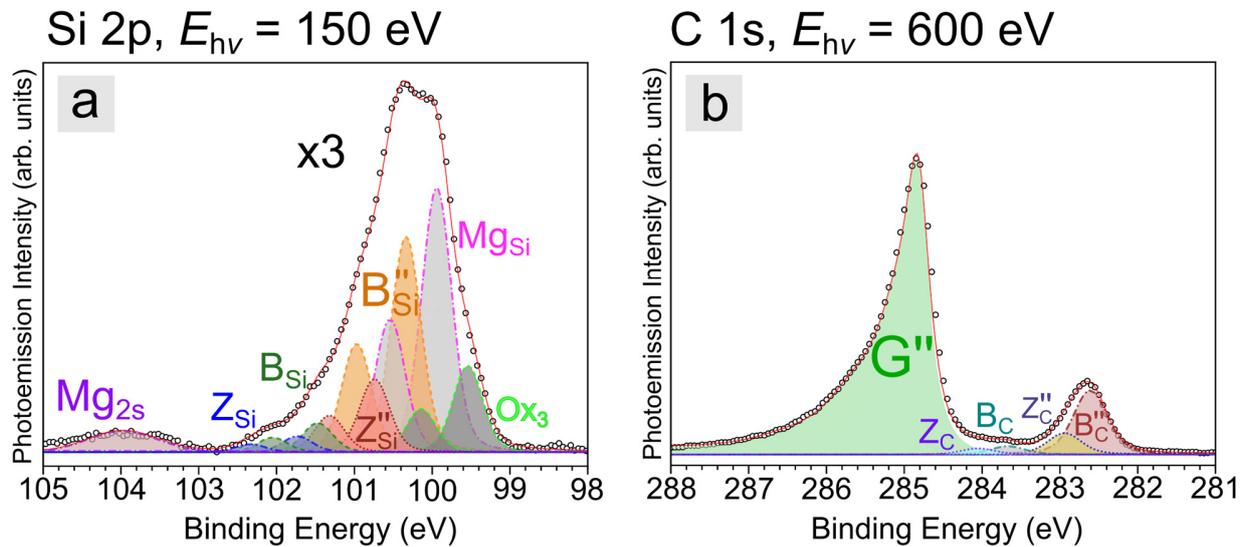

**Figure 9.** X-ray photoemission spectra showing the Si 2p and C 1s core levels of Mg-QFSBLG after ≈ 6 hours of ambient air exposure (and after annealing in UHV) at an x-ray energy of $E_{h\nu}$ = 150 eV and 600 eV, respectively. **(a)** Si 2p core level, y-scale magnified ×3 from Fig. 6a. **(b)** C 1s core level. Component labels in **a** are the same as Fig. 6c, d and labels in **b** are the same as Fig. 7c, d. Component $Mg_{Si}$ (and $Ox_3$) in **a**, has decreased in intensity relative to the other components causing the overall spectrum to change shape from Fig. 6c, d. Component $B_C''$ has not shifted in binding energy.

**Scanning Tunneling Microscopy of Mg-Intercalated graphene on SiC(0001).** A second EMLG sample was intercalated with Mg in UHV, then exposed to air for approximately 45 minutes during transfer to the low temperature UHV STM. Prior to scanning, the sample was annealed in UHV (see Methods). Figure 10 shows the STM topography map of the Mg-intercalated sample surface. The surface of the Mg-intercalation is almost identical to the Ca-intercalation sample in Fig. 4c, and does not show the (6√3×6√3)R30° Moiré pattern characteristic of pristine EMLG (Fig. 4a).

This offers further support that Mg-QFSBLG is relatively air-stable. The Mg-QFSBLG sample (Fig. 10) also shows very similar post-intercalation bright ($5\times10^{13}$ cm$^{-2}$) and dark ($4\times10^{14}$ cm$^{-2}$) spots as seen in Figs. 4b and c, further supporting our claims that the dark and post-intercalation bright spots are generic to graphenes intercalated at the SiC/buffer layer interface.

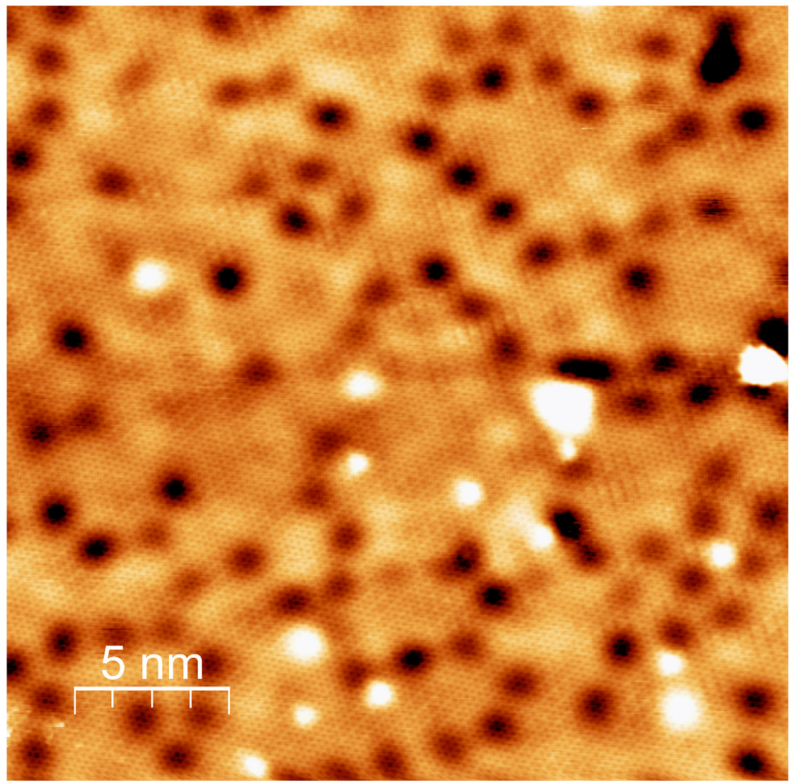

**Figure 10.** Scanning tunneling microscope (STM) micrograph of an Mg-QFSBLG sample exposed to air for ≈ 45 minutes before transferring to the STM (0.3 V, 200 pA, 77 K).

# Conclusion

In this communication we report on the structure of graphene on SiC(0001) intercalated with either Ca or Mg. By combining LEED, XPS, and STM with DFT modeling, we show that the favored route for Ca intercalation is insertion at the buffer layer – SiC(0001) interface, forming an ordered ($\sqrt{3}\times\sqrt{3}$)R30° reconstruction relative to the graphene and strong chemical interactions with the Si-termination of SiC. The intercalated Ca breaks the buffer layer-SiC covalent bonding, freestands the buffer layer to form Ca-QFSBLG and decreases the graphene workfunction, implying high n-type doping. Similarly, we find that the Mg intercalates EMLG samples at the buffer layer – SiC(0001) interface to form Mg-QFSBLG and also exhibits high n-type doping. Stronger chemical interaction between Si and Mg drives the formation of an ordered ($\sqrt{3}\times\sqrt{3}$)R30° reconstruction to the SiC. Unlike Ca-QFSBLG, Mg-QFSBLG exhibits remarkable air stability even after ≈ 6 hours of ambient air exposure. This work shows the first successful (to our knowledge) Mg-intercalation of graphene on SiC(0001) to form a low work function (n-type doped) and highly air stable graphene, which may find potential applications in highly conductive and transparent electrodes.

Furthermore, we see no evidence of Ca or Mg intercalation between the graphene layers under the conditions for intercalation used here, consistent with the predictions of our DFT modeling. While we cannot rule out that further processing could result in additional intercalation between graphene layers, our results already prompt a re-examination of previous studies where it was assumed that Ca intercalated *only* between graphene layers and *not* at the buffer layer – SiC interface.[8, 12, 19-20, 25]

## Methods

**Samples and Intercalation Procedure.** Two types of graphene samples were used in the experiments – epitaxial monolayer graphene synthesized on semi-insulating 6H-SiC(0001) (EMLG)[36] and hydrogen intercalated bilayer graphene synthesized on semi-insulating 6H-SiC(0001) (H-QFSBLG).[38] H-QFSBLG results are discussed in Supporting Information Section 1.3. The 6H-SiC(0001) semi-insulating substrates nominally on-axis (≈ 0.1 deg offcut) were obtained from II-VI Incorporated. Prior to graphene synthesis, the 6H-SiC(0001) substrates were etched in-situ (in an Aixtron/Epigress VP508 horizontal hot wall reactor) with high purity $H_2$ (200 mbar) at ≈ 1843 K. Epitaxial monolayer graphene was then synthesized in ≈ 20 minutes on these substrates while holding the temperature at ≈ 1843 K in an Ar atmosphere (100 mbar). This results in a sample with majority monolayer coverage on the SiC(0001) terraces, as well as regions of bilayer and small amounts of trilayer (see Supporting Information Figure S27 for Raman spectroscopic characterization of layer number for a sister sample). To fabricate H-QFSBLG, $H_2$ was flowed at 80 slm (900 mbar) for 15 – 75 minutes upon cooling to 1223 – 1423 K.[36] Large area Hall measurements on batch samples of this type show sheet density and carrier concentration vary by 10 – 15 %. Prior to measurement and intercalation, all samples were cleaned *via* annealing in UHV (≈ 1 × $10^{-10}$ mbar) between 673 and 773 K (determined with a single-color pyrometer *and* thermocouple) for approximately 8 hours (for XPS/LEED/SECO measurements). All alkaline earth metal depositions were undertaken while the samples were at room temperature.

The intercalation procedure at the Australian Synchrotron (in which the LEED, XPS and SECO measurements were taken) is as follows. For Ca-intercalation, samples (≈ 4 mm × 4 mm) were mounted on the same sample holder, consisting of 2 EMLG and 2 H-QFSBLG samples. Approximately 6 nm of Ca was deposited (as judged by a quartz crystal microbalance, QCM) using a WEZ effusion cell (MBE Komponenten) loaded with Ca metal (dendritic pieces, 99.99%, Sigma-Aldrich) in a pyrolytic boron nitride (PBN) crucible. The deposition required heating of the crucible to ≈ 698 K and exposure of 15 minutes to the Ca flux. After deposition, the sample was heated between ≈ 753 K for 1.5 hours to intercalate and clean the sample surface. This procedure constituted the 1st intercalation step. This intercalation procedure was then repeated for the 2nd intercalation step, except the anneal temperature was ≈ 736 K. All intercalation step data is shown Supporting Information Section 1.2/2.2, whereas in the manuscript we show the results for only the pristine samples and final intercalation step.

For the Mg-intercalation, the samples were mounted similarly (1 EMLG, 1 H-QFSBLG), and an NTEZ effusion cell (MBE Komponenten) was used with a PBN crucible loaded with Mg (1/8-inch turnings, 99.95%, Sigma-Aldrich). In the first intercalation step, ≈ 7.5 nm was deposited onto the samples (cell temperature 673 K, 10 minutes of Mg flux exposure) which were then heated to ≈ 773 K for ≈ 2 hours. Immediately afterwards, ≈ 3.8 nm (cell temperature 673 K, 5 minutes Mg flux exposure) of Mg was deposited before heating the sample to ≈ 773 K for ≈ 2 hours. These steps constituted the 1st intercalation step. In the 2nd intercalation step, the effusion cell was heated to 673 K, and the sample exposed to 10 minutes of Mg flux before annealing at 623 K for 1.5 hours. This sample was then exposed to ambient air for ≈ 6 hours and annealed at ≈ 623 K for 1.5 hours.

The Mg-QFSBLG sample in Fig. 8b was a sister sample that was similarly intercalated at the Australian Synchrotron, but this time in a 3-step Mg-intercalation procedure with the final Mg-intercalation yielding similar results to the final (*i.e.* 2$^{nd}$) Mg-intercalation of the above sample. In the first Mg-intercalation step, ≈ 15 nm of Mg was deposited on the sample (cell temperature 673 K, 20 minutes of Mg flux exposure), which was then heated to 463 K for ≈ 1.5 hours. For the 2$^{nd}$ Mg-intercalation, ≈ 15 nm of Mg was deposited on the sample (cell temperature 673 K, 20 minutes of Mg flux exposure), which was then heated to 623 K for ≈ 1.5 hours. For the 3$^{rd}$ Mg-intercalation, ≈ 15 nm of Mg was deposited on the sample (cell temperature 673 K, 20 minutes of Mg flux exposure), which was then heated to 623 K for ≈ 1.5 hours. See Supporting Information Figures S28 and S29 for the C 1s and Si 2p core levels, respectively.

The Mg-QFSBLG sample in Fig. 10 was fabricated via the intercalation of an EMLG sample. The same evaporator (NTEZ, PBN cell) and Mg source were used in a different UHV chamber. For this sample, the cell was heated to a temperature of 673 K and the sample was exposed to the Mg flux for 25 minutes (≈ 18.8 nm) before heating to 623 K for 30 minutes. The sample was then taken from storage in the UHV chamber after ≈ 1 month and mounted onto an STM holder in ambient (≈ 45 minutes) before being loaded into the STM chamber. The sample was cleaned at ≈ 673 K for ≈ 2 hours before scanning.

The Ca-intercalated graphene sample studied by STM (Fig. 4) was prepared as follows. An EMLG sample was cleaned in the CreaTec STM UHV (≈ 1×10$^{-10}$ mbar) chamber by annealing at ≈ 743 K for 1.2 hours (all temperatures for the STM were determined with a single-color pyrometer). The depositions were carried out with a CreaTec effusion cell using a PBN crucible loaded with Ca (dendritic pieces, 99.99 %, Sigma-Aldrich) that was baked at 473 K overnight prior to depositions. In the 1$^{st}$ intercalation (Fig. 4b), ≈ 1 nm of Ca was deposited while the sample was held at room temperature (all thicknesses were determined with a CreaTec QCM). After deposition, the sample was heated to a temperature of 653 K, held for 10 minutes, after which the sample was ramped to ≈ 743 K and held at this temperature for 1.5 hours to facilitate intercalation and remove excess surface Ca. For the 2$^{nd}$ Ca-intercalation, another 1 nm of Ca was deposited after re-calibration with the QCM. After deposition, the sample was heated to ≈ 653 K and held for 30 minutes, then ramped to ≈ 743 K and held for 1 hour before allowing the sample to cool naturally.

**Low Energy Electron Diffraction (LEED).** LEED measurements were conducted at the Australian Synchrotron on the Soft X-ray beamline using an 8-inch LEED spectrometer (OCI Vacuum Microengineering Inc.). Initial measurement coordinates on pristine samples were recorded, and for each subsequent intercalation / deposition, the same measurement positions were recorded. All measurements were taken at an energy of 100 eV. In Fig. 1a we show an example of a pristine EMLG, and in Fig. 1b and c shows a sister sample (grown in the same growth run) that was intercalated (hence the rotation of the sample). The LEED images in Fig. 1 are color inverted in order to enhance visibility.

**X-ray Photoelectron Spectroscopy (XPS).** XPS data was taken at the soft x-ray beamline at the Australian Synchrotron,[88] equipped with a PHOIBOS 150 (9 channeltrons) detector. The pass energy of the analyzer was set at 10 eV for the O 1s core levels (energy step of 0.1 eV – see Supporting Information Sections 1.5 and 2.4), and 5 eV for all other core levels (energy step of 0.05 eV). Each recorded spectrum was energy calibrated using the Au $4f_{7/2}$ line at 84.0 eV from an Au foil in electrical contact with the sample. Inelastic contributions were removed from each XPS spectrum in the main manuscript by subtraction of a Shirley line shape (except for Ca 2p spectra – see Supporting Information Section 1.5). The Si 2p spectra were obtained at $E_{h\nu}$ = 600 eV (bulk sensitive) and 150 eV (surface sensitive) – and were fit with Voigt functions with a Lorentzian full width at half maximum (FWHM) of 0.1 eV and Gaussian FWHM of 0.4 or 0.5 eV. The Lorentzian FWHM, which was kept constant, was determined from pristine QFSBLG samples, as these samples were relatively pure and acted as standards in the experiment. Additionally, the area ratio of the Si 2p spin-orbit split peaks (and Mg 2p/Ca 2p core levels) was strictly kept as $2p_{1/2}:2p_{3/2}$ = 1:2.[89] The spin-orbit splitting of the Si 2p core level after averaging and taking the standard deviation of all values was found to be 0.61 ± 0.02 eV, in excellent agreement with previously measured values.[89-90]

The C 1s 'graphene' peak (un-intercalated) was fit with a Breit-Wigner-Fano (BWF) asymmetric lineshape[66] (a type of asymmetric Lorentzian, see Supporting Information Section 1.1 for further details); as this lineshape was found to fit component G more accurately than the often used Doniach-Šunjić lineshape.[67] Upon intercalation, the doped graphene peak component G' was phenomenologically fit using a model that incorporated a BWF lineshape convoluted with 3 Voigt functions (the details of which can be found in Supporting Information Sections 1.1,1.2 and 2.2), to approximate the plasmon effects expected for a highly doped graphene lineshape which was recently described.[69-70] All C 1s core level SiC-related components were fit with Voigt lineshapes using a Gaussian width of 0.4 eV and Lorentzian width of 0.2 eV[49] (resulting in a FWHM = 0.52 eV, all parameters are outlined in Table S3 and Table S12 in Supporting Information Section 1.1 and 2.1, respectively). Care was taken to keep the Gaussian and Lorentzian contributions constant, *i.e.* the width of the Voigt functions did not change for the same components. Additionally, stringent conditions were placed on the binding energy location of the fitted peaks, which typically did not vary more than the uncertainty of the measurement, taken as ± 0.05 eV. The position of each component was calculated by averaging the available measurements (including the 1st intercalation step which was omitted in the main text – see Supporting Information, Sections 1.1 and 2.1) of the same component (in the same sample) across the various x-ray energies ($E_{h\nu}$). If the standard deviation was higher than the uncertainty in the measurement (*i.e.* > 0.05 eV), then the standard deviation was taken as the uncertainty in the measurement.

To qualitatively assess the surface sensitivity of the fitted components, the relative intensity (*RI*) was calculated by using the bulk SiC components of EMLG – $B_{Si}/B_C$ as the intensity reference (*i.e.* the RI of $B_{Si}$ and $B_C$ = 1). The *RI* ratio (*RI$_R$*) was then used as a qualitative metric of the surface sensitivity of a particular component by division of the lower $E_{h\nu}$ *RI* with the *RI* of the same component at higher $E_{h\nu}$. In this regime, an *RI$_R$* > 1 signifies a more surface-like component, and an *RI$_R$* ≤ 1 signifies a more bulk-like component.

**Secondary Electron Cut-Off (SECO).** SECO data (Fig. 8) was taken at $E_{h\nu}$ = 100 eV to find the workfunction ($W_f$) of the clean and intercalated graphene samples. The pass energy of the analyzer was set at 2 eV. A bias of –9 V was applied to the sample in order to effectively measure the low energy cutoff, and all samples were measured while perpendicular to the analyzer – as this is known to increase the accuracy of $W_f$ measurement.[91] The $W_f$ was then determined by the intersection of 2 linear fits to the steep edge and background, as is common with SECO determinations of the $W_f$.[92-93]

**Scanning Tunneling Microscopy (STM)**. All STM images were taken with a CreaTec low temperature STM. Micrographs shown in Figs. 4a, b were taken at 77 K, whereas the micrograph in Fig. 4c was taken at 4.6 K. In Fig. 4, a cut Pt-Ir tip was used, whereas an etched W tip was used in Fig. 10. Micrographs were analyzed using WSxM.[94]

**Density Functional Theory (DFT)**. We use density functional theory (DFT) calculations as implemented in the Vienna ab initio Simulation Package (VASP) to calculate the intercalation energy of Ca on SiC and graphene[95]. The electron exchange and correlation is described using the Perdew-Burke-Ernzehof (PBE) form of the generalized gradient approximation (GGA)[96]. A semi-empirical functional (DFT-D2) is employed to describe van der Waals interactions in the system[97]. The kinetic energy cutoff for the plane-wave basis set is set to 500 eV. We use a 9 × 9 × 1 Γ-centered k-point mesh for sampling the Brillouin zone. In all cases, the SiC/graphene system is modeled by a supercell comprised of three layers of 3×3 SiC crystal along with 2√3×2√3 graphene layers. The graphene is under a tensile strain of 7.6 %. The intercalation energy ($E_I$) of Ca is calculated by:

$$E_I = E(SiC/graphene + Ca) - E(SiC/graphene) - E(Ca),$$

where $E$(SiC/graphene), $E$(Ca) and $E$(SiC/graphene+Ca) are the energy of SiC/graphene heterostructure, atomic energy of Ca in its bulk state, and the energy of SiC/graphene system upon Ca intercalation, respectively. Here, a negative number signifies that the coordination is thermodynamically/energetically favorable, whereas a positive number signifies that the coordination is not energetically favorable.

AUTHOR INFORMATION

**Corresponding Authors**

*Email (J. C. Kotsakidis): jimmy.kotsakidis@monash.edu

*Email (A. L. Vazquez de Parga): al.vazquezdeparga@uam.es

*Email (M. S. Fuhrer): michael.fuhrer@monash.edu

**Author Contributions**

J.C.K. and A.L.V.P. performed the first Ca intercalations in the STM, from which a method was developed and guided future intercalation experiments at the Australian Synchrotron. J.C.K. suggested performing the Mg experiments, and developed the method for Mg-intercalation based upon the Ca-intercalation recipe. J.C.K. wrote the experimental proposal (including the methods) for the Australian Synchrotron experiments with necessary assistance from A.T. and M.S.F., and some grammatical corrections from A.G.C.. The data at the Australian Synchrotron was collected by J.C.K., A.G.C., A.L.V.P., A.T., C.L. and M.E.. Ca and Mg crucibles were first loaded at Monash University and transported to the Australian Synchrotron by J.C.K. (under vacuum), and mounted with help from A.T.. All starting samples were grown by D.K.G, R.M-W., M.D. and K.D., and M.C. post-processed the samples. S.P. and J.C.K. performed lab-based XPS, later analyzed by J.C.K.. J.C.K. performed Raman spectroscopy measurements with the aid of D.K.G and R.M-W, and was analyzed by J.C.K.. Y.Y. and N.M. performed the DFT calculations. J.C.K. composed the manuscript and all figures, with intellectual contributions from D.K.G. and M.S.F. The manuscript was written through contributions given by all authors. All authors have given approval to the final version of the manuscript.

**Funding Sources**

J.C.K. acknowledges the Australian Government Research Training Program and the Monash Centre for Atomically Thin Materials (MCATM) for financial support. J.C.K. and M.S.F. acknowledge funding support from the Australian Research Council (ARC) Laureate Fellowship (FL120100038). J.C.K., M.S.F., N.M and Y.Y. also acknowledge the ARC Centre of Excellence in Future Low-Energy Electronics (CE170100039) for financial support. N.M. and Y.Y also gratefully acknowledge computational support from the Monash Campus Cluster, NCI computational facility and Pawsey Supercomputing Facility. A.L.V.P. acknowledges funding support from the Ministerio de Ciencia Innovatión y Universidades project PGC2018-093291-B-I00 and Comunidad de Madrid project NMAT2D-CM P2018/NMT-4511. D.K.G., R.M-W., M.D., K.M.D., S.P. and M.C. acknowledges support by core programs at the U.S. Naval Research Laboratory funded by the Office of Naval Research. This research was undertaken (in part) on the soft x-ray beamline at the Australian Synchrotron, part of ANSTO.

**Notes**

The authors declare no competing financial interests.

**ACKNOWLEDGMENT**

J.C.K. is pleased to acknowledge Glenn Jernigan (U.S. Naval Research Laboratory, NRL) for assistance with initial XPS data interpretation, experimental insight, and helpful discussions regarding preliminary data and Iolanda Di Bernardo (Monash University) for helpful discussions regarding XPS data. J.C.K. also acknowledges the hospitality of NRL, at which some of the follow up experimental investigations were performed.

**Associated Content**

Elucidation of all Ca- and Mg-intercalation steps of EMLG, along with detailed tables outlining parameters of the component fits including the omitted 1$^{st}$ intercalation steps. Ca-intercalation of H-QFSBLG – LEED, XPS (C 1s and Si 2p core level spectra) and SECO measurements are shown. Measured C 1s graphene lineshapes for Ca- and Mg-intercalated EMLG (i.e Ca- and Mg-QFSBLG) are compared to theoretical lineshapes for highly doped graphene. The O 1s and Ca 2p XPS spectra for Ca-intercalated EMLG and H-QFSBLG are shown. STM micrographs of Ca deposition on top of EMLG are shown, as well as method for cleaning the graphene surface of excess Ca. Detailed STM micrographs of the dark and post-intercalation bright features are shown.

C 1s and Si 2p core level spectra of the attempted Mg-intercalation of H-QFSBLG is shown. The O 1s and Mg 2p core level spectra of Mg-intercalated EMLG, and Mg 2p spectra of H-QFSBLG are shown. The effect of long-term ambient exposure of Mg-QFSBLG is shown *via* lab-based C 1s/ Si 2p XPS spectra and Raman mapping spectra. The Si 2p and C 1s core levels of Mg-intercalated sample in Fig. 8b is shown.

**ABBREVIATIONS**

Ca-QFSBLG, Ca-intercalated quasi-freestanding bilayer graphene; $E_B$, binding energy; $E_{h\nu}$, x-ray energy; EMLG, epitaxial monolayer graphene; H-QFSBLG, H-intercalated quasi-freestanding bilayer graphene; LEED, low energy electron diffraction; Mg-QFSBLG, Mg-intercalated quasi-freestanding bilayer graphene; SECO, secondary electron cut-off; STM, Scanning Tunneling Microscopy; QFSBLG, quasi-freestanding bilayer graphene; XPS, X-ray photoelectron spectroscopy.

**FOR TABLE OF CONTENTS ONLY**

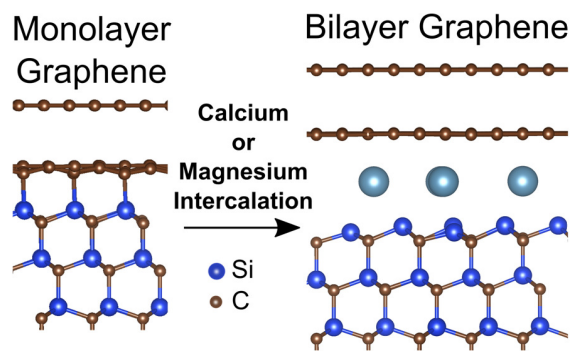